\documentclass[12pt,a4paper]{article}
\usepackage{epsfig,amsmath,amssymb}
\usepackage[colorlinks=true,linktocpage=true,linkcolor=blue,citecolor=blue]{hyperref}
\usepackage{cite}
\usepackage{graphicx}
\usepackage{dcolumn}
\usepackage{bm}
\usepackage{subfig}
\usepackage{float}
\usepackage{tabularx}
\usepackage{appendix}
\usepackage{hhline}
\usepackage{color}
\usepackage{slashed}
\usepackage{caption}
\begin{document}
\begin{flushright}
{\normalsize
}
\end{flushright}
\vskip 0.1in
\begin{center}
{\large {\bf Dissociation of heavy quarkonia in a weak magnetic field}}
\end{center}
\vskip 0.1in
\begin{center}
Mujeeb Hasan$^\dag$\footnote{mhasan@ph.iitr.ac.in} and Binoy Krishna Patra$^\dag$
\footnote{binoy@ph.iitr.ac.in}
\vskip 0.02in
{\small {\it $^\dag$ Department of Physics, Indian Institute of
Technology Roorkee, Roorkee 247 667, India}}
\end{center}
\vskip 0.01in
\addtolength{\baselineskip}{0.4\baselineskip} 

\begin{abstract}
We examined the effects of the weak magnetic field 
on the properties of heavy quarkonia immersed in a
thermal medium of quarks and gluons and studied how 
the magnetic field affects the quasi-free dissociation
of quarkonia in the aforementioned medium. For that 
purpose, we have revisited the general 
structure of gluon self-energy tensor in the presence 
of a weak magnetic field in thermal medium and obtained 
the relevant structure functions using the imaginary-time 
formalism. The structure functions give rise to the real and 
imaginary parts of the resummed gluon propagator, which 
further give the real and imaginary parts of the dielectric 
permittivity. The real and imaginary parts of the dielectric 
permittivity will be used to evaluate the real and imaginary 
parts of the complex heavy quark potential. We have 
observed that the real-part of the potential is found to be 
more screened, whereas the magnitude of the imaginary-part of 
the potential gets increased on increasing the value of 
both temperature and magnetic field. In addition to this, 
we have observed that the real-part gets slightly more screened 
while the imaginary part gets increased in the presence of 
a weak magnetic field as compared to their counterparts in 
the absence of a magnetic field (pure thermal). The increase 
in the screening of the real-part of the potential leads
to the decrease of binding energies of $J/\Psi$ and $\Upsilon$,
whereas the increase in the magnitude of the imaginary part
leads to the increase of thermal width with the 
temperature and magnetic field both. Also the 
binding energy and thermal width in the presence of 
weak magnetic field become 
smaller and larger, respectively, as compared to 
that in the pure thermal case. With the observations 
of binding energy and thermal width in hands, we have 
finally obtained the dissociation temperatures for 
$J/\Psi$ and $\Upsilon$, which become slightly lower in 
the presence of weak magnetic field. 
{\em For example}, with $eB = 0m_\pi^2$ the $J/\psi$ and 
$\Upsilon$ are dissociated at $1.80T_c$ and $3.50T_c$, 
respectively, whereas with $eB = 0.5m_\pi^2$ they 
dissociated at slightly lower value $1.74T_c$ and 
$3.43T_c$, respectively. This observation leads 
to the slightly early dissociation of quarkonia 
because of the presence of a weak magnetic field.
\end{abstract}

\vspace{1mm}

\section{Introduction}
Lattice QCD predicted that at sufficiently high temperatures
and/or densities the quarks and gluons confined inside
hadrons get deconfined into a medium of quarks and gluons coined
as quark-gluon Plasma. In the last few decades a large number
of experiment has been involved in identifying this new state 
of matter in ultrarelativistic heavy-ion collisions (URHICs) at
RHIC and LHC.  However, for the noncentral events in URHICs, 
a strong 
magnetic field is generated at the very early 
stages of the collisions due to very high relative velocities 
of the spectator quarks with respect to the fireball. 
Depending on the centralities of the collisions, the 
strength of the magnetic fields may vary from $m_{\pi}^2$ 
($\sim 10^{18}$ Gauss) at RHIC to 10 $m_{\pi}^2$ at 
LHC~\cite{Skokov:IJMPA24'2009,Voronyuk:PRC83'2011}. 
Motivated by this, in the recent past many 
theoretical works have started emerging to explore 
the effects of this strong magnetic field on the various  
QCD phenomena~\cite{Fukushima:PRD78'2008,
Braguta:PRD89'2014,Kharzeev:PRL106'2011,Gusynin:PRL73'1994}.
Earlier the nascent strong magnetic field was thought to decay 
very fast with time, resulting the magnetic field of weaker 
strength. However, it was 
later found that the realistic 
estimates of electrical conductivity of the medium 
may elongate the life-time of the magnetic field
~\cite{Tuchin:AHEP2013'2013,Mclerran:NPA929'2014,
Rath:PRD100'2019}. It thus becomes imperative to investigate
the effects of both strong and weak magnetic field on the 
signature of the novel matter produced in URHICs.

The heavy quarkonia is one of the 
probe to study the properties of nuclear matter under
extreme condition of temperature and magnetic field, because 
the heavy quark pairs are formed in URHICs on a very short time-scale 
$\sim 1/2m_Q$ (where $m_Q$ is the mass of 
the charm or bottom quark), which is similar to the 
time-scale at which the magnetic field is generated. 
Therefore the study of the effects of magnetic field on the
properties of heavy quarkonia is worth of investigation.
We have recently studied the properties of quarkonia 
in strong magnetic field. However, as we know the quarkonia, 
the physical resonances of $Q \bar Q$ states, are formed in the 
plasma frame at a time, $t_F$ (=$\gamma \tau_F$), 
which is order of $1-2fm$ depending on the resonances and 
their momenta. By the time elapsed, the magnetic field may 
become weak, so in our present study, we aim to understand 
theoretically the properties of heavy quarkonia and their 
dissociation in the presence of weak magnetic field
($T^2>|q_fB|$, $T^2>m_f^2$, where $|q_f|$ ($m_f$) is the 
absolute electric charge (mass) of 
the $f$-th quark flavour). As we 
know that, in order to study the dissociation of quarkonia 
the perturbative computation of heavy quarkonium potential 
is needed. 

Our understanding of heavy quarkonium has taken a major step forward in computing effective field theories (EFT) from the underlying theory - QCD, such as non-relativistic QCD (NRQCD)\cite{Bodwin:PRD51'1995} and potential NRQCD~\cite{Brambilla:NPB566'2000}, which
are synthesized successively by separating the intrinsic scales of heavy quark bound states (e.g. mass, velocity, binding energy) as well as the thermal medium-related scales (e.g. $T$, $gT$, $g^2 T$) in the weak-coupling system, in overall comparison with $\Lambda_{QCD}$. However, in the relativistic collisions that are created at URHICs, the separation of scales in an EFT is not always apparent, meaning it is often difficult to construct a potential model. An alternative approach is a  first-principle lattice QCD simulation in which one studies spectral functions derived from Euclidean meson 
correlation~\cite{Alberico:PRD77'2008}. The construction of spectral functions, however, is problematic
because the temporal range at large temperatures decreases.
For this reason studies of quarkonia using finite temperature potential models are useful as a complement to lattice studies.
The perturbative computations of the potential at 
high temperatures show that the potential of 
$Q \bar Q$ is complex~\cite{Laine:JHEP03'2007 }, 
where the real part is screened due to the existence 
of deconfined color charges~\cite{Matsui:PLB178'1986 } 
and the imaginary part ~\cite{Beraudo:NPA806'2008 } 
assigns the thermal width to the resonance. Therefore 
the physics of quarkonium dissociation in a medium 
has been refined in the last two decades, where the 
resonances were initially thought to be dissociated 
when the screening is strong enough, {\em i.e.} the 
real-part of the potential is too weak to keep 
the $Q\bar Q$ pair together. Nowadays, the dissociation 
is thought to be  primarily because of the widening of 
the resonance width arising either from the 
inelastic parton scattering mechanism mediated 
by the spacelike gluons, known as 
Landau damping~\cite{Laine:JHEP03'2007 } or 
from the gluo-dissociation process during which
color singlet state undergoes into a color octet 
state by a hard thermal gluon
~\cite{Brambilla:JHEP1305'2013}. The latter processes 
take precedence when the medium temperature
is lower than the binding energy of the particular 
resonance. This dissociates the quarkonium even at 
lower temperatures where the probability of color 
screening is negligible.
Recently one of us estimated the imaginary-part of the potential 
perturbatively, where the inclusion of a 
confining string term makes the (magnitude) imaginary component 
smaller~\cite{Lata:PRD89'2014,Lata:PRD88'2013}, compared to
the medium modification of the perturbative term alone
\cite{Adiran:PRD79'2009}. Gauge-gravity duality also indicates 
that in strong coupling limit the potential also develops an 
imaginary component beyond a critical separation of 
$Q \bar Q$ pair~\cite{Binoy:PRD92'2015,Binoy:PRD91'2015}.
Moreover lattice studies have also shown that the potential 
may have a sizable imaginary part~\cite{Rothkopf:PRL'2012}.
There are, however, other processes which may cause the 
depopulation of the resonance states either through the 
transition from ground state to the excited states during 
the non adiabatic evolution of quarkonia~\cite{Bagchi:MPLA30'2015} 
or through the swelling or shrinking of states due to the Brownian 
motion of $Q \bar Q$ states in the parton 
plasma~\cite{Binoy:NPA708'2002}. Very recently the change in the 
properties of heavy quarkonia immersed in a weakly-coupled
thermal QCD medium has been described by HTL permittivity
~\cite{Lafferty:arxiv:1906.00035}. They used the generalized 
Gauss law in conjunction to linear response theory to obtain 
the real and imaginary parts of the heavy quark potential, 
where a logarithmic divergence in imaginary part is found 
due to string contribution at large $r$.
They have circumvented by regularizing weak infrared
diverging ($1/p$) term in the resummed gluon propagator 
by choosing the regulation scale in terms of Debye mass.
There is another recent work~\cite{Guo:PRD100'2019}, where 
a nonperturbative term induced by the dimension two gluon 
condensate besides the usual HTL resummed contribution is 
included in the resummed gluon propagator to obtain 
the string contribution in the potential, in 
addition to the Karsch Mehr Satz (KMS) potential~\cite{KMS}.

The abovementioned studies are 
attributed for a thermal medium in the absence of a 
magnetic field. However, as mentioned earlier that 
a magnetic field is also 
generated in the heavy ion collisions, thus the influence of 
a homogeneous and constant external magnetic field 
on the heavy meson spectroscopy has been investigated 
quantum mechanically subjected to a three-dimensional harmonic 
potential and Cornell potential plus spin-spin 
interaction term~\cite{Alford:PRD88'2013,Bonati:PRD92'2015}.
Further, the effect of a constant uniform magnetic field 
on the static quarkonium potential at zero and finite 
temperature~\cite{Bonati:PRD94'2016} and on the screening 
masses~\cite{Bonati:PRD95'2017} have been investigated. 
The momentum diffusion coefficients of heavy quarks 
in a strong magnetic field along the directions parallel 
and perpendicular to the magnetic field at the leading order 
in QCD coupling constant has been 
studied~\cite{Fukushima:PRD93'2016}. Recently 
we have explored the effects of strong magnetic 
field on the properties of the heavy-quarkonium in 
finite temperature by computing the real part of 
the $Q \bar Q$ potential~\cite{Mujeeb:EPJC77'2017} 
in the framework of perturbative thermal QCD and 
studied the dissociation of heavy quarkonia due to 
the color screening. Successively, we made an attempt 
to study the dissociation of heavy quarkonia 
due to Landau damping in presence
of strong magnetic field by calculating the real
and imaginary parts of the heavy quark potential in
presence of strong magnetic field~\cite{Mujeeb:NPA995'2020}.  
The complex heavy quark potential in presence of strong 
magnetic field has also been obtained in~\cite{Balbeer:PRD97'2018}. Very recently we have 
also investigated the strong magnetic field-induced 
anisotropic interaction in heavy quark bound 
states~\cite{Salman:2004.08868}. The effects 
of strong magnetic field on the wakes in the induced charge density and in 
the potential due to the passage of highly energetic 
partons through a thermal QCD medium has also been investigated~\cite{Mujeeb:1901.03497}. 
Recently, the dispersion 
spectra of a gluon in hot QCD medium 
in presence of strong as well as weak magnetic field limit 
is studied~\cite{karmakar:EPJC79'2019}. The effect of the 
strong magnetic field on the collisional energy loss of heavy quark moving in a magnetized thermal partonic medium has 
been studied~\cite{Balbeer:arxiv2002.04922}. Also the anisotropic momentum diffusion and the drag coefficients of heavy quarks have been computed in a strongly magnetized quark-gluon plasma beyond the static limit within the framework of Langevindynamics~\cite{Balbeer:arxiv2004.11092}.

In the present study, we aim to obtain the complex heavy quark 
anti-quark potential in an environment of temperature and weak 
magnetic field. For that purpose, we first start with the
evaluation of gluon self energy in the similar environment
using the imaginary-time formalism.
As the quark-loop is only affected with the magnetic field 
thus, the quark-loop in the said environment 
is now dictated by both the scales namely the 
magnetic field as well as  the 
temperature, whereas for the gluon-loop, the temperature 
is the only available scale in the medium as the 
gluon-loop is not affected with the magnetic field. 
Furthermore, we have revisited the general structure of 
gluon self energy tensor in presence of weak magnetic field in 
thermal medium and obtained the relevant structure functions. 
Hence the real and imaginary parts of the resummed gluon 
propagator have been obtained, which give the real and imaginary 
parts of the dielectric permittivity. The real and imaginary 
parts of the dielectric permittivity will inturn give the real 
and imaginary parts of the complex heavy quark potential. 
The real part of the potential is 
used in the Schr\"{o}dinger equation to obtain the 
binding energy of heavy quarkonia whereas the imaginary 
part is used to calculate the thermal width. Finally, 
we have obtained the dissociation temperatures 
of heavy quarkonia and studied how the dissociation
temperatures get affected in presence of magnetic field. 

Thus, our work proceeds as follows. In section 2, we will
calculate the gluon self energy in a weak magnetic field 
wherein, we will discuss the general structure of gluon 
self energy and resummed gluon propagator at finite 
temperature in presence of weak magnetic field and will
calculate the relevant form factors in subsection 2.1 
and subsection 2.2, respectively. Thus, the real and 
imaginary parts of the resummed gluon propagator will 
give the real and imaginary parts of the dielectric 
permittivity in subsection 3.1, which gives the 
real and imaginary parts of complex heavy quark potential 
in subsection 3.2. We will use the real and imaginary 
parts of the potential to obtain the binding energy and 
thermal width in subsection 4.1 and 4.2, respectively, 
which will then give the dissociation temperatures 
of heavy quarkonia in subsection 4.3. Finally,
we will conclude our findings in section 5.

\section{Gluon self energy in a weak magnetic field}
In this section we will evaluate the gluon self energy in 
a weak magnetic field. As we know that for the evaluation
of gluon self energy, we need to evaluate both the quark
loop and gluon loop contributions in presence of 
weak magnetic field. Because of weak magnetic field,
only the quark loop will get affected whereas the gluon 
loop remain as such. Now, we will first start with the 
quark-loop contribution to gluon self energy
\begin{eqnarray}
i\Pi^{\mu\nu}_{ab}(Q)&=&-\int\frac{d^4K}{(2\pi)^4}Tr
\left[ i g t_b \gamma^\nu i S(K) i g t_a \gamma^\mu i S(P)
\right],
\nonumber\\
&=&\sum_f\frac{g^2\delta_{ab}}{2}\int\frac{d^4K}{(2\pi)^4}Tr
\left[ \gamma^\nu i S(K) \gamma^\mu i S(P)\right],
\label{self_energy}
\end{eqnarray}
where $P=(K-Q)$ and $Tr(t_a t_b)=\frac{\delta_{ab}}{2}$. 
The $S(k)$ is the quark propagator in a weak magnetic field
which can be written upto order of $O(q_fB)^2$ as
\cite{ayala:1805.07344} 
\begin{eqnarray}
iS(K)=i\frac{(\slashed{K}+m_f)}{K^2-m^2_f}-q_fB
\frac{\gamma_1\gamma_2(\slashed{K}_\parallel+m_f)}{(K^2-m_f^2)^2}
-2i(q_f B)^2\frac{[K_\perp^2(\slashed{K}_\parallel+m_f)
+\slashed{K}_\perp(m_f^2-K_\parallel^2)]}
{(K^2-m_f^2)^4},
\label{weak_propagator}
\end{eqnarray}
where $m_f$ and $q_f$ are the mass and charge of the 
$f^{th}$ flavor quark. According to the following 
choice of metric tensors,
\begin{eqnarray*}
g^{\mu\nu}_\parallel&=& {\rm diag} (1,0,0-1),\\
~g^{\mu\nu}_\perp&=&{\rm diag} (0,-1,-1,0),
\end{eqnarray*}
the four-momentum suitable in a magnetic field 
directed along the $z$ axis, 
$n^\mu =(0,0,0,-1)$, is given by
\begin{eqnarray}
K^{\mu}_\parallel&=&(k_0,0,0,k_z),\label{momentum_parallel}\\
K^{\mu}_\perp&=&(0,k_x,k_y,0),\label{momentum_perpendicular}\\
K_{\parallel}^2&=&k_{0}^2-k_{z}^2,\\ 
K_{\perp}^2&=&k_{x}^2+k_{y}^2.
\end{eqnarray}
The above Eq.\eqref{weak_propagator} can be recast in the 
following form 
\begin{eqnarray}
iS(K)=S_0(K)+S_1(K)+S_2(K), 
\label{weak_propagator1}
\end{eqnarray}
where 
$S_0(K)$ is the contribution of the order $O[(q_fB)^0]$,
$S_1(K)$ is the contribution of the order $O[(q_fB)^1]$
and $S_2(K)$ is the contribution of the order $O[(q_fB)^2]$.
Using Eq.\eqref{weak_propagator1}, the Eq.\eqref{self_energy} 
can be written as
\begin{eqnarray}
\Pi^{\mu\nu}(Q)=-\sum_f\frac{ig^2}{2}\int\frac{d^4K}
{(2\pi)^4}Tr\left[\gamma^\nu \lbrace S_0(K)+S_1(K)+S_2(K)
\rbrace \gamma^\mu \lbrace S_0(P)+S_1(P)+S_2(P)\rbrace 
\right].
\label{self_energy1}
\end{eqnarray}
After simplifying, the above gluon self energy given by 
Eq.\eqref{self_energy1} can be expressed as follows
\begin{eqnarray}
\Pi^{\mu\nu}(Q)=\Pi^{\mu\nu}_{(0,0)}(Q)+\Pi^{\mu\nu}_{(1,1)}(Q)
+2\Pi^{\mu\nu}_{(2,0)}(Q)+O[(q_fB)^3],
\label{self_energy2}
\end{eqnarray}
where 
\begin{eqnarray}
\Pi^{\mu\nu}_{(0,0)}(Q)&=&-\sum_f\frac{ig^2}{2}\int\frac{d^4K}
{(2\pi)^4}Tr[\gamma^\nu S_0(K)\gamma^\mu S_0(P)],\label{pi_00}\\
\Pi^{\mu\nu}_{(1,1)}(Q)&=&-\sum_f\frac{ig^2}{2}\int\frac{d^4K}
{(2\pi)^4}Tr\lbrace \gamma^\nu S_1(K) \gamma^\mu S_1(P)
\rbrace,\label{pi_11}\\
\Pi^{\mu\nu}_{(2,0)}(Q)&=&-\sum_f\frac{ig^2}{2}\int\frac{d^4K}
{(2\pi)^4}Tr\left[\gamma^\nu S_2(K) \gamma^\mu S_0(P)\right].
\label{pi_20}
\end{eqnarray}
The term $\Pi^{\mu\nu}_{(0,0)}$ is of the order $O[(q_fB)^0]$, where 
$\Pi^{\mu\nu}_{(1,1)}$ and $\Pi^{\mu\nu}_{(2,0)}$ both are of the order 
$O[(q_fB)^2]$. The term which is of the order $O[(q_fB)^1]$ 
vanishes. Substituting the values of $S_0$, $S_1$ and $S_2$ 
in Eq.\eqref{pi_00}, 
Eq.\eqref{pi_11} and Eq.\eqref{pi_20} by comparing Eq.\eqref{weak_propagator} 
with Eq.\eqref{weak_propagator1},  we get 

\begin{eqnarray}
\Pi^{\mu\nu}_{(0,0)}(Q)&=&\sum_f\frac{ig^2}{2}\int\frac{d^4K}{(2\pi)^4}
\frac{Tr[\gamma^\nu(\slashed{K}+m_f)\gamma^\mu(\slashed{P}+m_f)]}
{(K^2-m^2_f)(P^2-m_f^2)},\nonumber\\
&=&\sum_f i2g^2\int\frac{d^4K}{(2\pi)^4}\frac{\left[P^\mu K^\nu+
K^\mu P^\nu-g^{\mu\nu}(K\cdot P-m_f^2)\right]}
{(K^2-m^2_f)(P^2-m_f^2)},\\
\Pi^{\mu\nu}_{(1,1)}(Q)&=&-\sum_f\frac{ig^2(q_fB)^2}{2}\int\frac{d^4K}
{(2\pi)^4}\frac{Tr[\gamma^\nu\gamma_1\gamma_2
(\slashed{K}_\parallel+m_f)\gamma^\mu\gamma_1\gamma_2
(\slashed{P}_\parallel+m_f)]}
{(K^2-m^2_f)^2(P^2-m_f^2)^2},\nonumber\\
&=&\sum_f 2ig^2(q_fB)^2\int\frac{d^4K}{(2\pi)^4}
\frac{\left[P_\parallel^\mu K_\parallel^\nu +K_\parallel^\mu 
P_\parallel^\nu +(g_\parallel^{\mu\nu}-g_\perp^{\mu\nu})
(m_f^2-K_\parallel\cdot P_\parallel)\right]}
{(K^2-m^2_f)^2(P^2-m_f^2)^2},\\
\Pi^{\mu\nu}_{(2,0)}(Q)&=&-\sum_f\frac{2ig^2(q_fB)^2}{2}\int\frac{d^4K}{(2\pi)^4}
\frac{Tr\left[\gamma^\nu\lbrace K_\perp^2
(\slashed{K}_\parallel+m_f)+\slashed{K}_\perp
(m_f^2-K_{\parallel}^2)\rbrace\gamma^\mu(\slashed{P}+m_f)\right]}
{(K^2-m_f^2)^4(P^2-m_f^2)},
\nonumber\\
&=&-\sum_f 4ig^2(q_fB)^2\int\frac{d^4K}{(2\pi)^4}
\left[\frac{M^{\mu\nu}}{(K^2-m_f^2)^4(P^2-m_f^2)}\right],
\end{eqnarray}
where
\begin{eqnarray}
M^{\mu\nu}&=&K_\perp^2\left[P^\mu K_\parallel^\nu+
K_\parallel^\mu P^\nu-g^{\mu\nu}(K_\parallel\cdot P-m_f^2)\right]
+(m_f^2-K_\parallel^2)\left[P^\mu K_\perp^\nu +K_\perp^\mu 
P^\nu-g^{\mu\nu}(K_\perp\cdot P)\right].~~~
\end{eqnarray}
Here the strong coupling $g$ runs with the magnetic field and 
temperature both, which is recently obtained in~\cite{ayala:PRD98'2018}
\begin{eqnarray}
\alpha_s(\Lambda^2,eB)=\frac{g^2}{4\pi}=\frac{\alpha_s(\Lambda^2)}{1+
b_1\alpha_s(\Lambda^2)\ln\left(\frac{\Lambda^2}
{\Lambda^2+eB}\right)},
\end{eqnarray}
with 
\begin{eqnarray}
\alpha_s(\Lambda^2)=\frac{1}{
b_1\ln\left(\frac{\Lambda^2}
{\Lambda_{\overline{MS}}^2}\right)},
\end{eqnarray}
where $\Lambda$ is set at $2\pi T$, $b_1=\frac{11N_c-2N_f}{12\pi}$ and
$\Lambda_{\overline{MS}}=0.176GeV$.

Before evaluating further, we will first discuss the structure
of gluon self energy in thermal medium in presence of weak
magnetic field in the next subsection.
 
\subsection{Structure of gluon self energy and resummed gluon propagator 
for thermal medium in the presence of weak magnetic field}
In this subsection, we will briefly discuss the general structure 
of gluon self energy tensor and resummed gluon propagator for thermal medium 
in the presence of weak
magnetic field. The general structure of gluon self energy in a 
thermal medium defined by the heat bath in local rest frame, 
$u^\mu=(1,0,0,0)$ and in the presence of magnetic field directed 
along the $z$-direction, $n_\mu=(0,0,0,-1)$ is recently obtained 
as follows~\cite{karmakar:EPJC79'2019} 
\begin{eqnarray}
\Pi^{\mu\nu}(Q)=b(Q)B^{\mu\nu}(Q)+c(Q)R^{\mu\nu}(Q)+d(Q)M^{\mu\nu}(Q)
+a(Q)N^{\mu\nu}(Q),
\label{self_decomposition}
\end{eqnarray}
where 
\begin{eqnarray}
B^{\mu\nu}(Q)&=&\frac{{\bar{u}}^\mu{\bar{u}}^\nu}{{\bar{u}}^2},\\
R^{\mu\nu}(Q)&=&g_{\perp}^{\mu\nu}-\frac{Q_{\perp}^{\mu}Q_{\perp}^{\nu}}
{Q_{\perp}^2},\\
M^{\mu\nu}(Q)&=&\frac{{\bar{n}}^\mu{\bar{n}}^\nu}{{\bar{n}}^2},\\
N^{\mu\nu}(Q)&=&\frac{{\bar{u}}^\mu{\bar{n}}^\nu+{\bar{u}}^\nu{\bar{n}}^\mu}
{\sqrt{{\bar{u}}^2}\sqrt{{\bar{n}}^2}},
\end{eqnarray}
the four vectors ${\bar{u}}^\mu$ and ${\bar{n}}^\mu$ used in the 
construction of above tensors are defined as follows
\begin{eqnarray}
\bar{u}^\mu &=&\left(g^{\mu\nu}-\frac{Q^\mu Q^\nu}{Q^2}\right)u_\nu,\\
\bar{n}^\mu &=&\left(\tilde{g}^{\mu\nu}-\frac{\tilde{Q}^\mu\tilde{Q}^\nu}
{\tilde{Q}^2}\right)n_\nu,
\end{eqnarray}
where ${\tilde{g}}^{\mu\nu}=g^{\mu\nu}-u^\mu u^\nu$ and 
$\tilde{Q}^\mu=Q^\mu-(Q.u)u^\mu$. Using the properties of 
projection tensors, the form factors appear in 
\eqref{self_decomposition} can be obtained as
\begin{eqnarray}
b(Q)&=&B^{\mu\nu}(Q)\Pi_{\mu\nu}(Q),
\label{form_b}\\
c(Q)&=&R^{\mu\nu}(Q)\Pi_{\mu\nu}(Q),
\label{form_c}\\
d(Q)&=&M^{\mu\nu}(Q)\Pi_{\mu\nu}(Q),
\label{form_d}\\
a(Q)&=&\frac{1}{2}N^{\mu\nu}(Q)\Pi_{\mu\nu}(Q)
\label{form_a}.
\end{eqnarray}
Now we can obtained the resummed gluon propagator in thermal medium
in presence of weak magnetic field. The general form of the resummed 
gluon propagator in Landau gauge can be written 
as~\cite{karmakar:EPJC79'2019} 
\begin{eqnarray}
D^{\mu\nu}(Q)=\frac{(Q^2-d)B^{\mu\nu}}{(Q^2-b)(Q^2-d)-a^2}
+\frac{R^{\mu\nu}}{Q^2-c}+\frac{(Q^2-b)M^{\mu\nu}}{(Q^2-b)(Q^2-d)-a^2}
+\frac{aN^{\mu\nu}}{(Q^2-b)(Q^2-d)-a^2}.
\end{eqnarray}
The point to be noted here is that, we required only the ``00''-component 
of resummed gluon propagator for deriving the heavy quark potential. 
Hence the ``00''-component of the propagator can be obtained as 
\begin{eqnarray}
D^{00}(Q)=\frac{(Q^2-d)\bar{u}^2}{(Q^2-b)(Q^2-d)-a^2},
\label{propagator_00}
\end{eqnarray}
where $R^{00}=M^{00}=N^{00}=0$. Now we will obtained the form factors
appear in the above propagator \eqref{propagator_00}. We will first
start with the form factor $a$, which can be obtained using Eq.\eqref{form_a}
with Eq.\eqref{self_energy2} as 
\begin{eqnarray}
a(Q)=a_0(Q)+a_2(Q),
\end{eqnarray}
where $a_0$ is of the order of $O(q_fB)^0$ and 
$a_2$ is of the order of $O(q_fB)^2$. An important point to be
noted here is that the zero magnetic field contribution of form 
factor $a$ vanishes, that is $a_0=0$, whereas $a_2$ gives the 
contribution of order $O(q_fB)^2$. However the contribution 
of form factor $a$ in the denominator of the 
propagator \eqref{propagator_00} appear as $a^2$, which becomes 
of the order of $O(q_fB)^4$. Since in the current theoretical 
calculation we are considering contribution upto $O(q_fB)^2$, 
so we can neglect the contribution appear from the form factor 
$a$. Thus, the ``00''-component of resummed gluon 
propagator upto $O(q_fB)^2$ can be written as
\begin{eqnarray}
D^{00}(Q)=\frac{\bar{u}^2}{(Q^2-b)},
\label{propagator_final}
\end{eqnarray}
so we end up with only one form factor $b$, which we will 
evaluate in the next subsection.
    
\subsection{Real and imaginary parts of the form factor $b(Q)$}
In this subsection, we will calculate the real and imaginary
parts of the form factor 
$b$. Using Eq.\eqref{form_b}, the form factor $b$ 
can be evaluated as follows 
\begin{eqnarray}
b(Q)&=&B_{\mu\nu}(Q)\Pi^{\mu\nu}(Q),\nonumber\\
b(Q)&=&\frac{{\bar{u}}_\mu{\bar{u}}_\nu}{{\bar{u}}^2}\Pi^{\mu\nu}(Q),\nonumber\\
&=&\left[\frac{u_\mu u_\nu}{{\bar{u}}^2}-\frac{(Q.u)u_\nu Q_\mu}{\bar{u}^2Q^2}-\frac{(Q.u)u_\mu Q_\nu}{\bar{u}^2Q^2}+\frac{(Q.u)^2Q_\nu Q_\mu}{\bar{u}^2Q^4}\right]\Pi^{\mu\nu}(Q),\nonumber\\
&=&\frac{u_\mu u_\nu}{{\bar{u}}^2}\Pi^{\mu\nu}(Q),
\label{correct_form}
\end{eqnarray}
where we have used transversality condition $Q_\mu\Pi^{\mu\nu}(Q)=Q_\nu\Pi^{\mu\nu}(Q)=0$, to arrive at Eq.
\eqref{correct_form}. Thus using Eq.\eqref{self_energy2}, the form factor b can be 
written upto $O[(q_fB)^2]$  as
\begin{eqnarray}
b(Q)=b_0(Q)+b_2(Q),
\label{formfactor_b}
\end{eqnarray}
where the form factors $b_0$ and $b_2$ are defined as follows
\begin{eqnarray}
b_0(Q)&=&\frac{u_\mu u_\nu}{\bar{u}^2}\Pi^{\mu\nu}_{(0,0)}(Q),
\label{formfactor_b0}\\
b_2(Q)&=&\frac{u_\mu u_\nu}{\bar{u}^2}[\Pi^{\mu\nu}_{(1,1)}(Q)+
2\Pi^{\mu\nu}_{(2,0)}(Q)].
\label{formfactor_b2}
\end{eqnarray}
\textbf{\underline{Form factor $b_0(Q)$ (order of $O[(q_fB)^0]$)}}:\\
\\
Here we will solve the form factor $b_0$. Using 
Eq.\eqref{formfactor_b0}, the form factor can be
written as
\begin{eqnarray}
b_0(Q)&=&\frac{u_\mu u_\nu}{\bar{u}^2}\Pi^{\mu\nu}_{(0,0)}(Q),
\nonumber\\
&=&\sum_f \frac{i2g^2}{\bar{u}^2}\int\frac{d^4K}{(2\pi)^4}\frac
{\left[2k_0^2-K^2+m_f^2\right]}
{(K^2-m^2_f)(P^2-m_f^2)}.
\end{eqnarray}
Now we will solve the form factor $b_0$ using the imaginary-time 
formalism, the detailed calculation for which has been shown 
in appendix~\ref{b_0}. 
Thus, the real and imaginary parts of the 
form factor $b_0$ in the static limit 
are given as follows
\begin{eqnarray}
{\rm Re}~b_0(q_0=0)&=&g^2T^2\frac{N_f}{6},\\
\left[\frac{{\rm Im}~b_0(q_0,q)}{q_0}\right]_
{q_0=0}&=&\frac{g^2T^2N_f}{6}\frac{\pi}{2q}.
\end{eqnarray} 
Now we will evaluate the gluonic contribution. The 
temporal component of gluon self energy due to the 
gluon-loop contribution can be calculated as
~\cite{Weldon:PRD26'1982,Pisarski:PRL63'1989},
\begin{eqnarray}
\Pi^{00}(q_0,q)=-g^2 T^2 \frac{N_c}{3}\left(\frac{q_0}
{2q}\ln\frac
{q_0+q+i\epsilon}{q_0-q+i\epsilon}-1\right)~,
\end{eqnarray}
which gives the real and imaginary parts of form 
factor $b_0$ due to the gluonic contribution in the
static limit
\begin{eqnarray}
{\rm Re}~b_0(q_0=0)&=&g^2T^2\left(\frac{N_c}{3}\right),\\
\left[\frac{{\rm Im}~b_0(q_0,q)}{q_0}\right]_{q_0=0}
&=&g^2T^2\left(\frac{N_c}{3}\right)\frac{\pi}{2q}.
\label{img_b0}
\end{eqnarray}
Now we add the quark and gluon-loop contributions together to
obtain the real and imaginary parts of form factor $b_0$ in the 
static limit as follows 
\begin{eqnarray}
{\rm Re}~b_0(q_0=0)&=&g^2T^2\left(\frac{N_c}{3}+\frac{N_f}{6}\right),
\label{real_b0}\\
\left[\frac{{\rm Im}~b_0(q_0,q)}{q_0}\right]_{q_0=0}
&=&g^2T^2\left(\frac{N_c}{3}+\frac{N_f}{6}\right)\frac{\pi}{2q}.
\label{img_b0}
\end{eqnarray}
Thus we can see that the form factor $b_0$ is independent of the
magnetic field as it is $O[(q_fB)^0]$ and depends only
on the temperature of the medium. This form factor $b_0$  
coincides with the HTL form factor $\Pi_L$ in absence of 
the magnetic field~\cite{Weldon:PRD26'1982,Pisarski:PRL63'1989}.

\textbf{\underline{Form factor $b_2(Q)$ (order of $O[(q_fB)^2]$)}}:\\
\\
Here we will discuss the form factor $b_2$, which is of the 
order of $O[(q_fB)^2]$. Using Eq.\eqref{formfactor_b2}, the 
form factor is given by 
\begin{eqnarray}
b_2(Q)&=&\frac{u_\mu u_\nu}{\bar{u}^2}[\Pi^{\mu\nu}_{(1,1)}(Q)+
2\Pi^{\mu\nu}_{(2,0)}(Q)],
\nonumber\\
&=&\sum_f \frac{i2g^2(q_fB)^2}{\bar{u}^2}\left[\int\frac{d^4K}{(2\pi)
^4}\left\lbrace\frac{\left(2k_0^2-K_\parallel^2+m_f^2\right)}
{(K^2-m^2_f)^2(P^2-m_f^2)^2}
-\frac{\left(8k_0^2K_\perp^2\right)}{(K^2-m^2_f)^4(P^2-m_f^2)}\right
\rbrace\right].~~~
\end{eqnarray}
We have calculated the real and imaginary parts of the 
form factor $b_2$ in the appendix~\ref{b_2}, which gives the 
real and imaginary parts of the form factor $b_2$ in the 
static limit as follows
\begin{eqnarray}
{\rm Re}~b_2(q_0=0)=\sum_f\frac{g^2}{12\pi^2 T^2}(q_fB)^2
\sum_{l=1}^{\infty}(-1)^{l+1}l^2K_0(\frac{m_fl}{T}).
\label{real_b2}
\end{eqnarray}
\begin{eqnarray}
\left[\frac{{\rm Im}~b_2(q_0,q)}{q_0}\right]_
{q_0=0}&=&\frac{1}{q}\left[\sum_f\frac{g^2(q_fB)^2}{16\pi T^2}
\sum_{l=1}^\infty(-1)^{l+1}l^2K_0\left(\frac{m_f l}{T}\right)
\right. \nonumber\\ &&\left. 
-\sum_f\frac{g^2(q_fB)^2}{96\pi T^2}
\sum_{l=1}^\infty(-1)^{l+1}l^2K_2\left(\frac{m_f l}{T}\right)
\right. \nonumber\\ &&\left.
+\sum_f\frac{g^2(q_fB)^2}{768\pi}\frac{(8T-7\pi m_f)}{m_f^2 T}
\right],
\label{img_b2}
\end{eqnarray}
where $K_0$ and $K_2$ are the modified Bessel functions 
of second kind.

\begin{figure}[t]
\begin{center}
\begin{tabular}{c c}
\includegraphics[width=7.6cm,height=8.4cm]{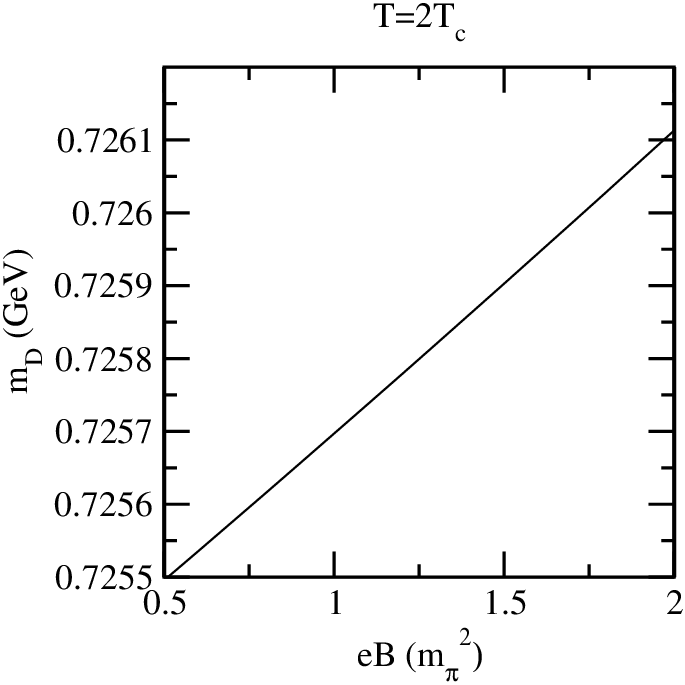}&~~~~~
\includegraphics[width=7.5cm,height=7.6cm]{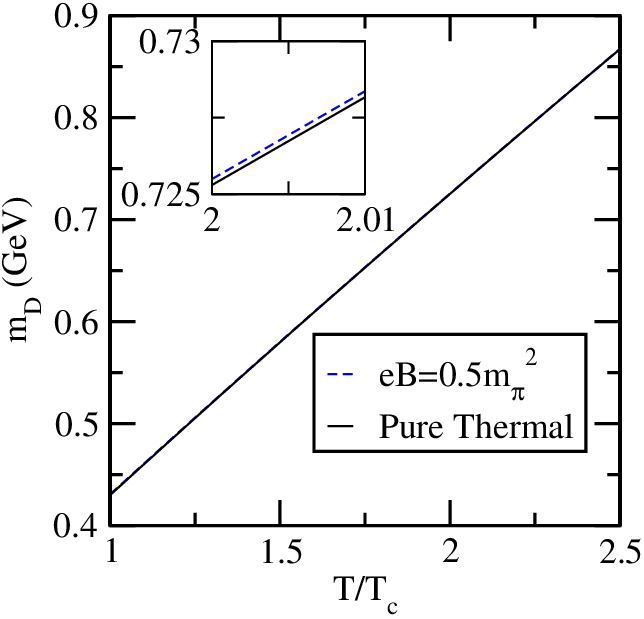}\\
\end{tabular}
\caption{Variation of Debye mass with magnetic field (left panel)
and with temperature (right panel).}
\label{debye}
\end{center}
\end{figure}

After obtaining the real and imaginary parts of
the form factor $b_0$ and $b_2$, we can write the 
real and imaginary parts of form factor $b$ using 
Eq.\eqref{formfactor_b} as follows
\begin{eqnarray}
{\rm Re}~b(q_0=0)&=&
g^2T^2\left(\frac{N_c}{3}+\frac{N_f}{6}\right)+
\sum_f\frac{g^2}{12\pi^2 T^2}(q_fB)^2
\sum_{l=1}^{\infty}(-1)^{l+1}l^2K_0(\frac{m_fl}{T}),
\label{real_b}\\
\left[\frac{{\rm Im}~b(q_0,q)}{q_0}\right]_
{q_0=0}&=&g^2T^2\left(\frac{N_c}{3}+\frac{N_f}{6}\right)\frac{\pi}{2q}
+\frac{1}{q}\left[\sum_f\frac{g^2(q_fB)^2}{16\pi T^2}
\sum_{l=1}^\infty(-1)^{l+1}l^2K_0\left(\frac{m_f l}{T}\right)
\right. \nonumber\\ &-&\left. 
\sum_f\frac{g^2(q_fB)^2}{96\pi T^2}
\sum_{l=1}^\infty(-1)^{l+1}l^2K_2\left(\frac{m_f l}{T}\right)
+\sum_f\frac{g^2(q_fB)^2}{768\pi}\frac{(8T-7\pi m_f)}{m_f^2 T}
\right],
~~\label{img_b}
\end{eqnarray}
where Eq.\eqref{real_b} is the real-part of the form factor 
in the static limit which gives the Debye screening mass in 
the presence of weak magnetic field as follows
\begin{eqnarray}
M_D^2=g^2T^2\left(\frac{N_c}{3}+\frac{N_f}{6}\right)+
\sum_f\frac{g^2}{12\pi^2 T^2}(q_fB)^2
\sum_{l=1}^{\infty}(-1)^{l+1}l^2K_0(\frac{m_fl}{T}).
\label{debyemass}
\end{eqnarray}
Thus, it is observed that Debye screening mass of the thermal 
medium in the presence of weak magnetic field is affected 
by both the temperature and magnetic field. Now in order to see
how the Debye mass is changed in the presence of weak magnetic field 
we have mentioned the leading order result of Debye mass 
for thermal medium in absence of magnetic 
(termed as ``Pure Thermal'')~\cite{Shuryak:ZETF'1978}.
\begin{eqnarray}
M^2_D({\rm Pure~Thermal})=g^2 T^2 
\left(\frac{N_c}{3} +\frac{N_f}{6}\right).
\end{eqnarray}
In the left panel of Fig.\ref{debye}, we have quantitatively
studied the variation of Debye mass with varying strength 
of weak magnetic field for a fixed value of temperature. We 
have observed that the debye mass is found to increase
with the varying strength of magnetic field. On the other hand, 
in the right panel
of Fig.\ref{debye}, we have studied the variation with the 
temperature for a fixed value of magnetic field and 
observed that the Debye mass is also found to increase 
with increasing temperature, but 
the increase of Debye mass with temperature is fast 
as compared to the slow increase with magnetic field. In
addition to this, we have also made a comparison of Debye
mass in presence of magnetic field with the one in absence 
of magnetic field and observed that the Debye mass in presence 
of weak magnetic field is found to be slightly higher as 
compared to the one in pure thermal case. 

\section{Medium modified heavy quark potential}
In this section we will discuss the medium modification to 
the potential between a heavy quark $Q$ and its anti-quark 
$\bar{Q}$ in the presence of weak magnetic field at 
finite temperature.
Since the mass of the heavy quark ($m_Q$) is very large, 
so the requirements - $m_Q \gg T \gg \Lambda_{QCD}$ and 
$m_Q \gg \sqrt{eB}$ are satisfied for the description of 
the interactions between a pair of heavy quark and 
anti-quark at finite temperature in a weak 
magnetic field in terms of quantum mechanical 
potential, that leads to the validity of 
taking the static heavy quark potential. 
Thus we can obtain the 
medium-modification to the vacuum potential in the 
presence of magnetic field by correcting both its 
short and long-distance part 
with a dielectric function $\epsilon(q)$ 
as 
\begin{equation}
V(r;T,B)=\int\frac{d^3q}{(2\pi)^{3/2}}
({e^{iq.r}-1})\frac{V(q)}{\epsilon(q)},
\label{pot_defn}
\end{equation}
where the $r$-independent term has subtracted to renormalize 
the heavy quark free energy, which is the perturbative 
free energy of quarkonium at infinite separation. The Fourier 
transform,  $V(q)$  of the perturbative part of the Cornell 
potential ($V(r)=-\frac{4\alpha_s}{3r}$) 
is given by
\begin{equation}
{V}(q)=-\frac{4}{3}\sqrt{\frac{2}{\pi}} 
\frac{\alpha_s}{q^2},
\label{ft_pot}
\end{equation}
and the dielectric permittivity, $\epsilon(q)$, 
embodies the effects of confined medium in the presence of 
magnetic field is to be calculated next. The important 
point to be noted here is that we have taken the Fourier 
transform of the perturbative part of the vacuum potential 
only, the reason for this is that we can not use the same 
screening scale for both Coulomb and string terms because 
of the non-perturbative 
nature of the string term. To include the non-perturbative 
part of the potential, we will use the method of dimension 
two gluon condensate. 
\subsection{The complex permittivity for a hot QCD medium 
in a weak magnetic field}
The complex dielectric permittivity, $\epsilon (q)$ is 
defined by the static limit of ``00''-component 
of resummed gluon propagator from the linear
response theory
\begin{equation}
\frac{1}{\epsilon (q)}=-\displaystyle
{\lim_{q_0 \rightarrow 0}}{q}^{2}D^{00}(q_{0}, 
q).
\label{dielectric}
\end{equation}
Now we will evaluate the ``00''-component of resummed 
gluon propagator. The real-part of the resummed 
gluon propagator in the static limit 
can be evaluated by using 
Eq.\eqref{propagator_final} and Eq.\eqref{real_b} 
\begin{eqnarray}
{\rm Re}~D^{00}(q_0=0)=\frac{-1}{q^2+M_D^2}.
\label{real_resummed}
\end{eqnarray}
The imaginary part of resummed
gluon propagator can be written in terms
of the real and imaginary parts of the 
form factor by using the following 
formula~\cite{Weldon:PRD42'1990}
\begin{eqnarray}
{\rm Im}~D^{00}(q_0,q)=\frac{2T}{q_0}\frac{{\rm Im}~b(q_0,q)}
{(Q^2-{\rm Re}~b(q_0,q))^2+({\rm Im}~b(q_0,q))^2},
\end{eqnarray}
which can be recast into the following form
\begin{eqnarray}
{\rm Im}~D^{00}(q_0,q)=2T\frac{\left[\frac{{\rm Im}~b(q_0,q)}
{q_0}\right]}
{(Q^2-{\rm Re}~b(q_0,q))^2+(q_0\left[\frac{{\rm Im}~b(q_0,q)}
{q_0}\right])^2},
\end{eqnarray}
in the static limit the above equation reduces to 
the simplified form
\begin{eqnarray}
{\rm Im}~D^{00}(q_0=0)=2T\frac{\left[\frac{{\rm Im}~
b(q_0,q)}{q_0}\right]_{q_0=0}}{(q^2+M_D^2)^2},
\label{resummed}
\end{eqnarray}
where we have substituted ${\rm Re}~b(q_0=0)=M_D^2$. 
Using Eq.\eqref{img_b} and the above Eq.\eqref{resummed}, 
the imaginary part of ``00''-component of resummed gluon
propagator can be written as follows 
\begin{eqnarray}
{\rm Im}~D^{00}(q_0=0,q)=\frac{\pi T M^2_{(T,B)}}{q(q^2+M_D^2)^2},
\label{img_resummed}
\end{eqnarray}
where we have defined the quantity $M^2_{(T,B)}$ as follows
\begin{eqnarray}
M_{(T,B)}^2&=&g^2T^2\left(\frac{N_c}{3}+\frac{N_f}{6}\right)
+\left[\sum_f\frac{g^2(q_fB)^2}{8\pi^2 T^2}
\sum_{l=1}^\infty(-1)^{l+1}l^2K_0\left(\frac{m_f l}{T}\right)
\right. \nonumber\\ &&-\left. 
\sum_f\frac{g^2(q_fB)^2}{48\pi^2 T^2}
\sum_{l=1}^\infty(-1)^{l+1}l^2K_2\left(\frac{m_f l}{T}\right)
+\sum_f\frac{g^2(q_fB)^2}{384\pi^2}\frac{(8T-7\pi m_f)}{m_f^2 T}
\right].~
\end{eqnarray}
Now we will obtain the real and imaginary parts 
of dielectric permittivity, before evaluating them
we will discuss the procedure to handle the 
nonperturbative part of the heavy quark potential.
The handling of the nonperturbative part of the potential is recently 
been discussed in~\cite{Guo:PRD100'2019}. The procedure 
is to include
a nonperturbative term in the real and imaginary parts
of the ``00''-component of resummed gluon propagator along
with the usual Hard Thermal Loop (HTL) propagator which 
we have obtained earlier. The real and imaginary parts of the 
nonperturbative (NP) term by using the dimension two gluon
condensate are given as follows
\begin{eqnarray}
{\rm Re}~D^{00}_{NP}(q_0=0,q)=-\frac{m_G^2}{(q^2+M_D^2)^2},\\
{\rm Im}~D^{00}_{NP}(q_0=0,q)=\frac{2\pi TM^2_{(T,B)}m_G^2}
{q(q^2+M_D^2)^3},
\end{eqnarray}
where $m_G^2$ is a dimensional constant, which can be related 
to the string tension through the relation 
$\sigma=\alpha m_G^2/2$. Thus, the real and 
imaginary parts of the ``00''-component of the 
resummed gluon propagator that consists of 
both the HTL and the NP 
contributions can be written as follows
\begin{eqnarray}
{\rm Re}~D^{00}(q_0=0,q)&=&-\frac{1}
{q^2+M_D^2}-\frac{m_G^2}
{(q^2+M_D^2)^2}
\label{real_propagator},\\
{\rm Im}~D^{00}(q_0=0,q)&=&\frac{\pi T M^2_{(T,B)}}
{q(q^2+M_D^2)^2}+\frac{2\pi T M^2_{(T,B)}m_G^2}
{q(q^2+M_D^2)^3}.
\label{imaginary_propagator}
\end{eqnarray}
Now substituting Eq.\eqref{real_propagator} and 
Eq.\eqref{imaginary_propagator} in Eq.\eqref{dielectric} 
gives the real and imaginary parts of the dielectric 
permittivity, respectively
\begin{eqnarray}
\frac{1}{{\rm Re}~\epsilon (q)}&=&\frac{q^2}
{q^2+M_D^2}+\frac{q^2 m_G^2}
{(q^2+M_D^2)^2}
\label{real_dielectric},\\
\frac{1}{{\rm Im}~\epsilon (q)}&=&-\frac{q\pi T M^2_{(T,B)}}
{(q^2+M_D^2)^2}-\frac{2q\pi T M^2_{(T,B)}m_G^2}
{(q^2+M_D^2)^3}.
\label{imaginary_dielectric}
\end{eqnarray}
We are now going to derive the real and imaginary 
parts of the complex potential from the real and imaginary 
parts of dielectric permittivities, respectively in 
the next subsection. The important point to be 
noted here is that the non perturbative terms in the
real and imaginary parts of the dielectric permittivity
will lead to the string contribution in the 
real and imaginary parts of the potential.

\subsection{Real and Imaginary parts of the potential}
\begin{figure}[h]
\begin{center}
\begin{tabular}{c c}
\includegraphics[width=7.5cm,height=7.5cm]{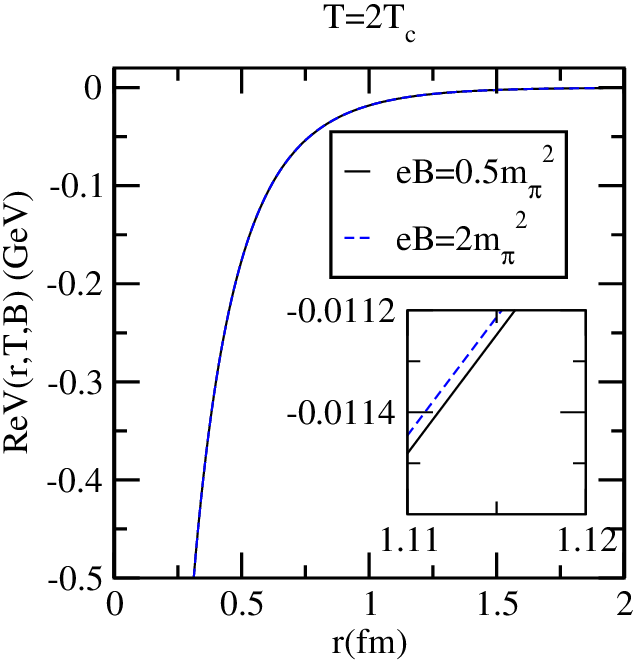}
&~~~~~
\includegraphics[width=7.5cm,height=7.5cm]{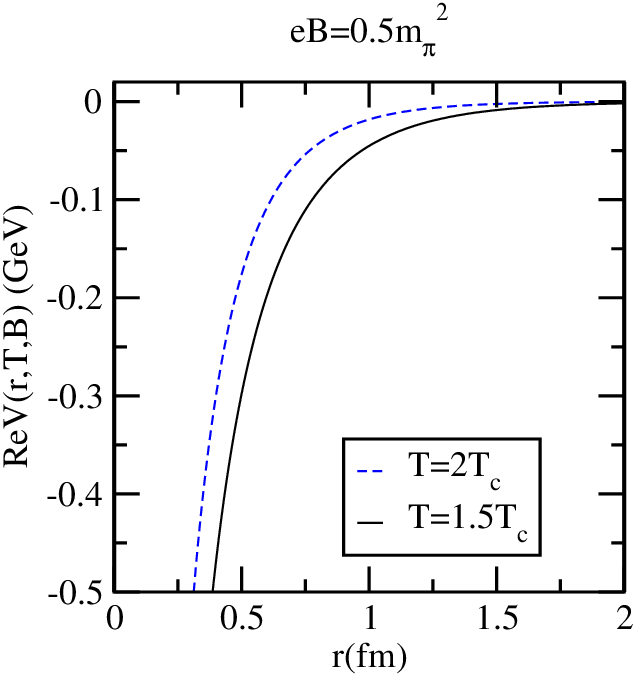}\\
\end{tabular}
\caption{Real-part of the potential for different strengths of 
magnetic field (left panel) and for different strengths of
temperature (right panel).}
\label{realb}
\end{center}
\end{figure}

Here we will calculate the real and imaginary parts 
of the heavy quark potential in presence of weak magnetic field.
The real-part of the dielectric permittivity 
in Eq.\eqref{real_dielectric} is substituted into the definition 
of potential in Eq.\eqref{pot_defn} to obtain the real-part of 
$Q \bar Q$ potential in the presence of weak magnetic 
field 
(with $\hat{r}=rM_{D}$)
\begin{eqnarray}
\rm{Re} V(r;T,B)&=&-\frac{4}{3}\alpha_s\left(\frac{e^{-\hat{r}}}
{r}+M_D\right)+\frac{4}{3}\frac{\sigma}{M_D}
\left(1-e^{-\hat{r}}\right),
\label{real_potential}
\end{eqnarray}
where the temperature and magnetic field dependence in the
potential enters through the 
Debye mass. While plotting the real-part of the 
potential we have excluded the non-local terms which are 
however, required to reduce the potential in the medium 
$V(r;T,B)$ to the vacuum potential in $(T, B) \rightarrow 0 $ 
limit. In Fig.\ref{realb}, we have plotted the real-part of the 
potential as a function of interquark distance ($r$). In the left
panel of Fig.\ref{realb}, we have plotted the real-part of the 
potential for different strengths of weak magnetic field like 
$eB=0.5m_\pi^2$ and $2m_\pi^2$ for a fixed value of temperature 
$T=2T_c$. We observed that on increasing 
the value of magnetic field the real-part become more screened.
Whereas in the right panel of Fig.\ref{realb}, the real-part
is plotted for different strengths of temperature like $T=1.5T_c$
and $T=2T_c$ and found to be more screened on increasing 
the value of temperature. Thus, the real-part of the potential
is found to be more screened on increasing the value of both 
temperature and magnetic field. This observation of the 
real-part of the potential can be understood in terms 
of the observation of the Debye mass which is found 
to be increased both with temperature and magnetic field 
as shown earlier in Fig.\ref{debye}.

\begin{figure}[h]
\begin{center}
\includegraphics[width=7.5cm,height=7.5cm]{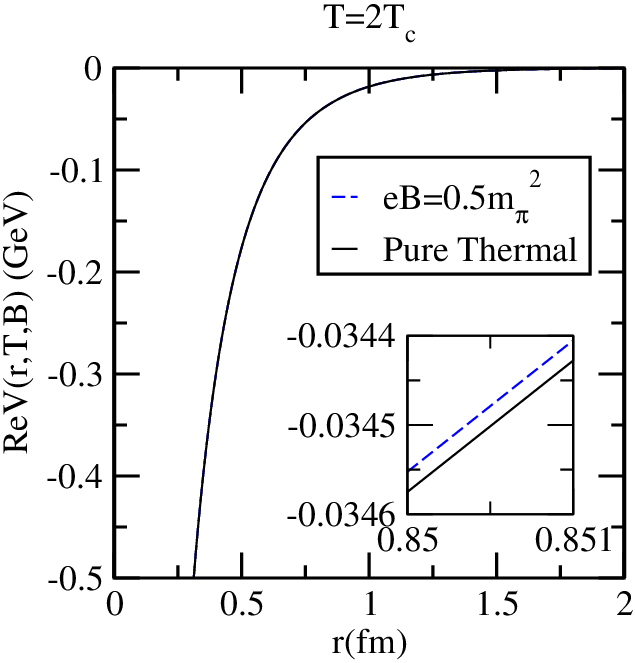}
\caption{Real-part of the potential in the absence and 
presence of weak magnetic field.}
\label{real_comp}
\end{center}
\end{figure}
We have made a comparison in 
Fig.\ref{real_comp} to see how the magnetic field 
will affect the real-part of the potential, for 
that we have plotted the real-part of the 
potential in presence of magnetic field with the 
one for pure thermal case. As we have 
seen in the right panel of Fig.\ref{debye} that the Debye 
mass in presence of magnetic field is slightly higher
as compared to the Debye mass in pure thermal medium,
that leads to the slightly more screening of the 
real-part of the potential in presence of weak magnetic 
field as compared to the same in the pure thermal case.   

We will now evaluate the imaginary-part of the
potential in presence of weak magnetic field. The 
imaginary-part of the potential is obtained by substituting 
the imaginary part of dielectric permittivity from 
Eq.\eqref{imaginary_dielectric} into the 
definition of the potential Eq.\eqref{pot_defn}
\begin{eqnarray}
\rm{Im} V_C(r;T,B)&=&-\frac{4}{3}\frac{\alpha_s T M^2_{(T,B)}}{M_D^2}\phi_2(\hat{r}),\\
\rm{Im} V_S(r;T,B)&=&-\frac{4 \sigma T M^2_{(T,B)}}{M_D^4}
\phi_3(\hat{r}),
\label{imaginary_potential}
\end{eqnarray}
where the function $\phi_2(\hat{r})$ and $\phi_3(\hat{r})$
are given in~\cite{Guo:PRD100'2019}
\begin{eqnarray}
\phi_2(\hat{r})&=&2\int_0^{\infty}\frac{zdz}{(z^2+1)^2}
\left[1-\frac{\sin z\hat{r}}{z\hat{r}}\right],\\
\phi_3(\hat{r})&=&2\int_0^{\infty}\frac{zdz}{(z^2+1)^3}
\left[1-\frac{\sin z\hat{r}}{z\hat{r}}\right],
\end{eqnarray}
and in the small $\hat{r}$ limit $(\hat{r}\ll 1)$, the above functions 
become 
\begin{eqnarray}
\phi_2(\hat{r})&\approx &-\frac{1}{9}{\hat{r}}^2\left(3\ln \hat{r}-
4+3\gamma_E\right),\\
\phi_3(\hat{r})&\approx&\frac{{\hat{r}}^2}{12}+\frac{{\hat{r}}^4}{900}
\left(15\ln\hat{r}-23+15\gamma_E\right).
\end{eqnarray}
It is worth mentioning that we considered the imaginary part of the potential within the
small distance limit ($\hat{r}=rM_D\ll 1$), so that it can be viewed as a perturbation. This could be relevant for the bound states of very heavy quarks, where Bohr radii, $r_B$ (=$\frac{n^2}{g^2 m_Q}$) of quarkonia are smaller than the Debye length, $\frac{1}{M_D}$. As we know that the former ($r_B$) is related to the scales of nonrelativistic heavy quark bound states in vacuum ($T=0$) 
and the scales associated to the thermal medium. In fact, the above condition ($r_B < \frac{1}{M_D}$) is translated to the hierarchy for the validity of potential approach ($m_Q> T~ {\rm or}~ gT$).

\begin{figure}[t]
\begin{center}
\begin{tabular}{c c}
\includegraphics[width=7.5cm,height=7.5cm]{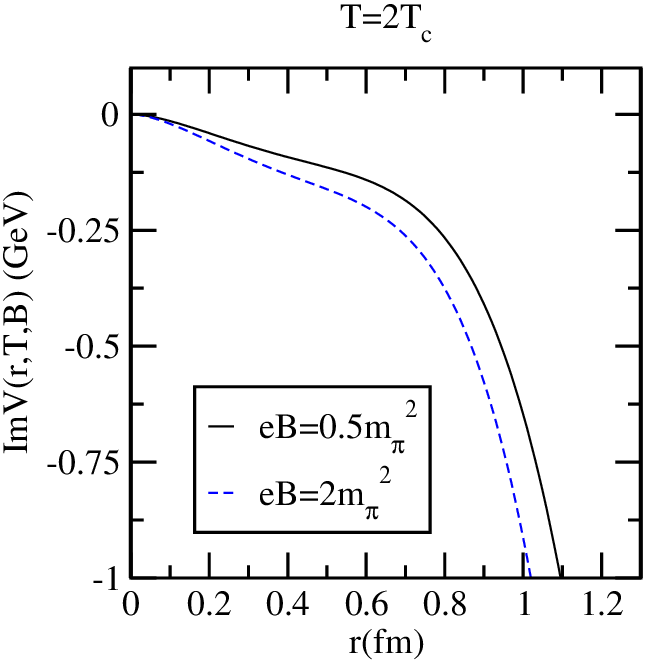}
&~~~~~\includegraphics[width=7.5cm,height=7.5cm]{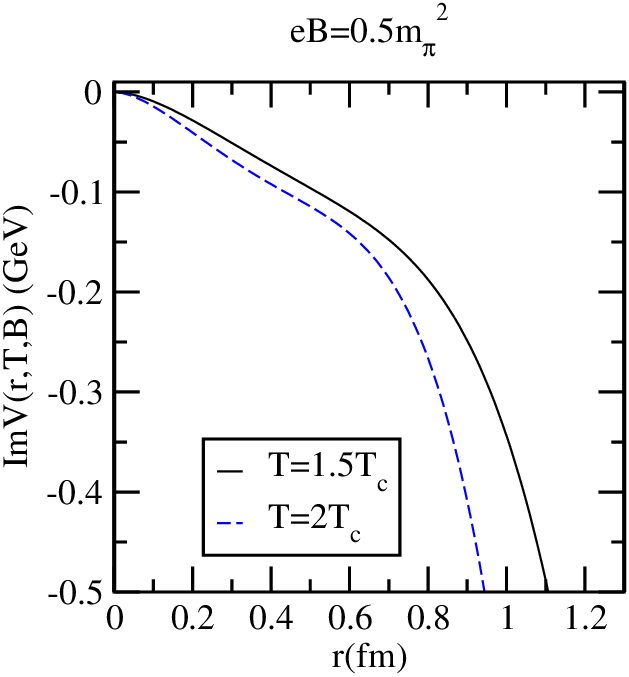}\\
\end{tabular}
\caption{Imaginary-part of the potential for different strengths of 
magnetic field (left panel) and for different strengths of
temperature (right panel).}
\label{imgb}
\end{center}
\end{figure}

\begin{figure}[h]
\begin{center}
\includegraphics[width=7.5cm,height=7.5cm]{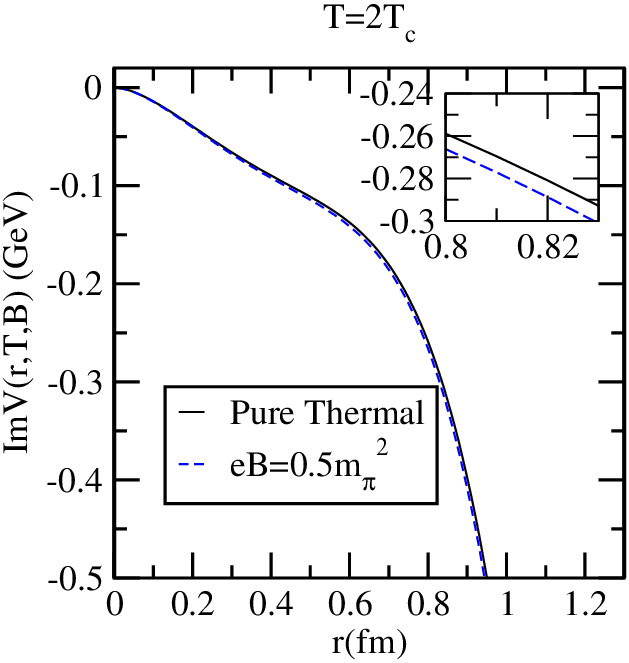}
\caption{Imaginary-part of the potential in the absence and 
presence of weak magnetic field.}
\label{img_comp}
\end{center}
\end{figure}

Similar to the real-part of the potential we have plotted the 
imaginary-part of the potential as a function 
of interquark distance ($r$) in Fig.\ref{imgb}. We have calculated 
the imaginary-part of the potential for different strengths of 
weak magnetic field like $eB=0.5m_\pi^2$ and $2m_\pi^2$ in the left
panel of Fig.\ref{imgb}. We found that on increasing 
the value of magnetic field the magnitude of imaginary-part 
gets increased. On the other hand, in the right panel of 
Fig.\ref{imgb}, the imaginary-part is calculated for different 
strengths of temperature like $T=1.5T_c$
and $T=2T_c$, here also the imaginary-part is found to increase 
with the temperature. Hence the magnitude of the imaginary-part of 
the potential gets increased with the value of temperature 
and magnetic field both. This observation also attributed to the 
fact that the Debye mass is found to be increased with temperature 
and magnetic field both.
Here also we have calculated the imaginary-part of the potential 
in presence of magnetic field with the one for pure thermal 
case in Fig.\ref{img_comp}, where we observed that the 
imaginary-part of the potential in presence of magnetic field 
is increased slightly as compared to the one in pure thermal 
case. 
\newpage

\section{Properties of quarkonia}
In this section we first explore the effects of weak magnetic 
field on the properties of heavy quarkonia. The obtained real
and imaginary parts of the heavy quark potential will be 
used to evaluate the binding energy and thermal width of 
the heavy quarkonia, respectively.

\subsection{Binding energy}
\begin{figure}[t]
\begin{center}
\begin{tabular}{c c}
\includegraphics[width=7.5cm,height=7cm]{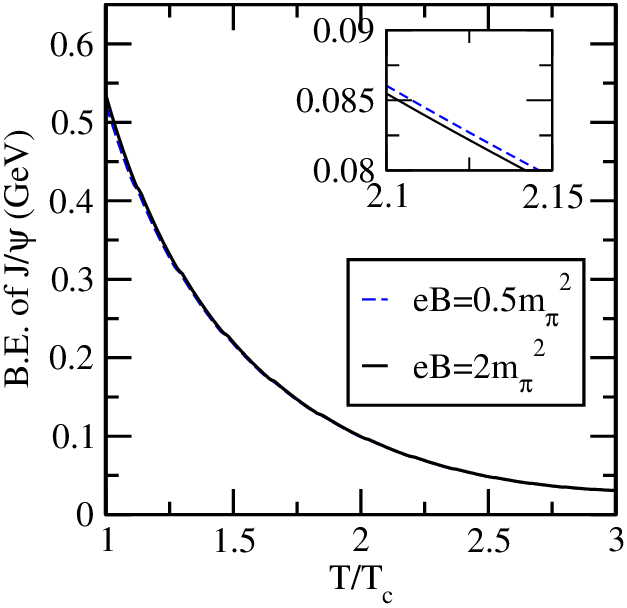}&
~~~~~\includegraphics[width=7.5cm,height=7cm]{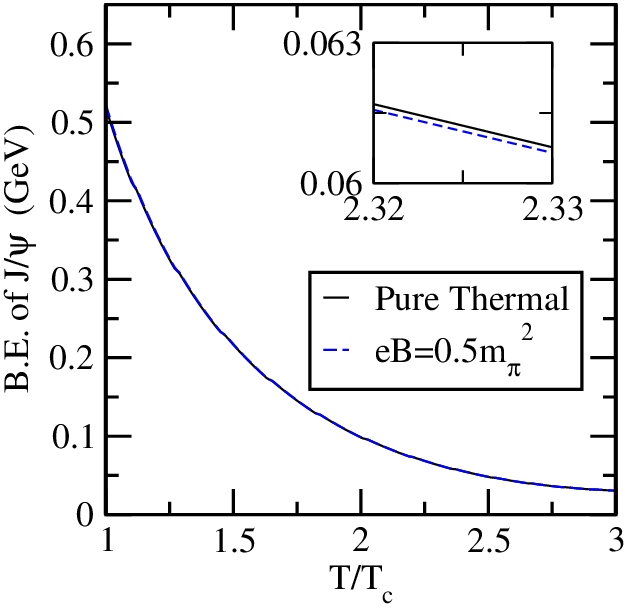}
\end{tabular}
\caption{The binding energy of $J/\psi$ as a function of 
temperature.} 
\label{psi}
\end{center}
\end{figure}

In this subsection, we have obtained the binding energy 
of $J/\psi$ and 
$\Upsilon$. In order to calculate the 
binding energy, the real part of 
the potential Eq.\eqref{real_potential} is put into the radial 
part of the Schr\"{o}dinger equation, which is then solved 
numerically to obtain the energy eigenvalues that inturns give 
the binding energies of quarkonia. To see how the presence of 
weak magnetic 
field affects the binding of quarkonia, we have plotted the 
binding energies of $J/\psi$ as a function of $T/T_c$ for 
different strengths of magnetic field in the left panel 
of Fig.\ref{psi}. We observed that the binding energy is 
found to decrease with the temperature and magnetic field
both, we can attribute this finding in terms of the increasing 
of screening with the temperature and magnetic field that 
we have observed in the real-part of the potential. The point 
to be noted here is that the difference between the values 
of binding energies plotted for the magnetic field $eB=0.5m_\pi^2$ 
and $eB=2m_\pi^2$ is pronounced at higher
temperature, this is in accordance with the validity of our 
work in the weak field limit $(T^2>|q_fB|)$. 
\begin{figure}[h]
\begin{center}
\begin{tabular}{c c}
\includegraphics[width=7.5cm,height=7cm]{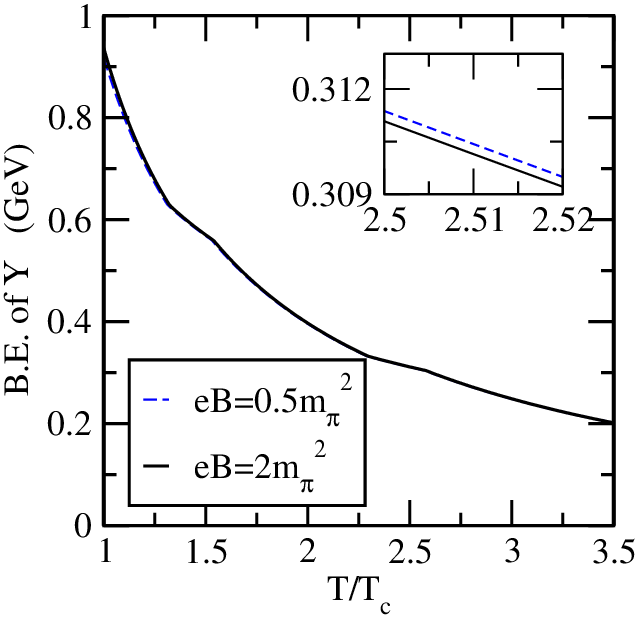}&
~~~~~\includegraphics[width=7.5cm,height=7cm]{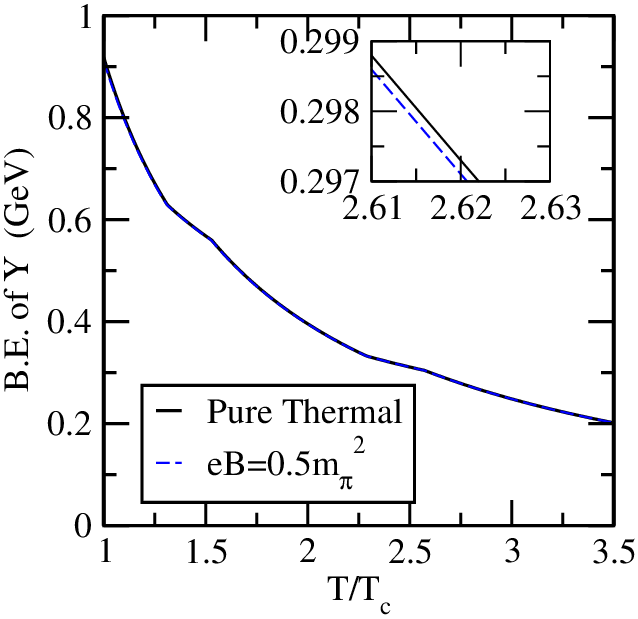}
\end{tabular}
\caption{The binding energy of $\Upsilon$ as a function of temperature.}
\label{upsilon}
\end{center}
\end{figure} 
  
In the right panel of Fig.\ref{psi}, we have also compared the 
binding energy of $J/\psi$ in presence of weak magnetic field with
the pure thermal case. We found that the binding energy in presence 
of magnetic field is smaller as compared to the one in 
pure thermal case, this is because the real-part of the potential
in presence of magnetic becomes more screened as compared to 
pure thermal case. The similar observation has also been
observed for $\Upsilon$, except that the value of binding 
energy for $\Upsilon$ is higher as compared to the value for 
$J/\Psi$. The variation of binding energy for $\Upsilon$ is 
studied in the left and right panel of Fig.\ref{upsilon}.  
\newpage

\subsection{Thermal width}
We will now use the imaginary part of the potential obtained in presence 
of weak magnetic field to estimate the broadening of the resonance states 
in a thermal medium. So
using the first-order time-independent perturbation theory, the width 
($\Gamma$) has been evaluated by folding with ($\Phi(r)$),
\begin{eqnarray}
\Gamma({\rm T,B})=-2\int_0^\infty \rm{Im}~V(r;T,B) |\Phi(r)|^2 d\tau,
\label{gammaT}
\end{eqnarray}
the wave function $\Phi(r)$ is taken to be the Coloumbic wave function 
for the ground state
\begin{eqnarray}
\Phi(r)=\frac{1}{\sqrt{\pi a_0^3}}e^{-r/a_0},
\end{eqnarray}
where $a_0$ is the Bohr radius of the 
heavy quarkonium system. 
Here we have used the imaginary-part of the potential
as a perturbation to obtain the thermal width, and for that 
purpose we have obtained the imaginary-part of the potential
in the small distance limit.   
 \begin{figure}[t]
\begin{center}
\begin{tabular}{c c}
\includegraphics[width=7.5cm,height=7.5cm]{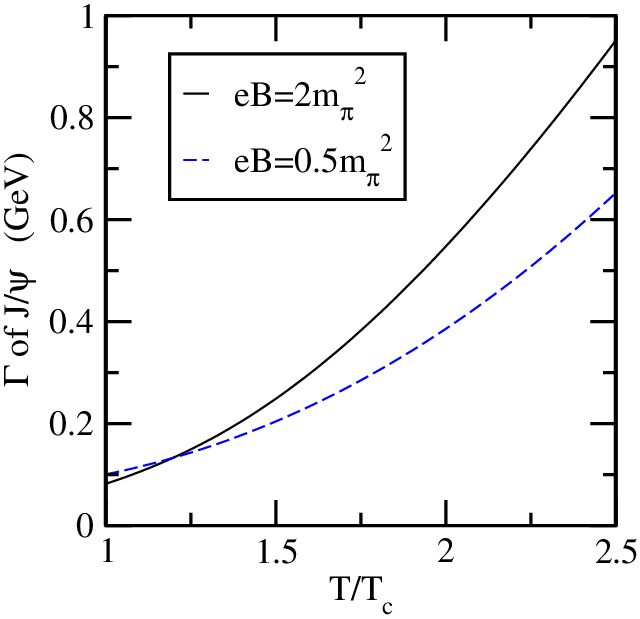}&
~~~~\includegraphics[width=7.5cm,height=7.5cm]{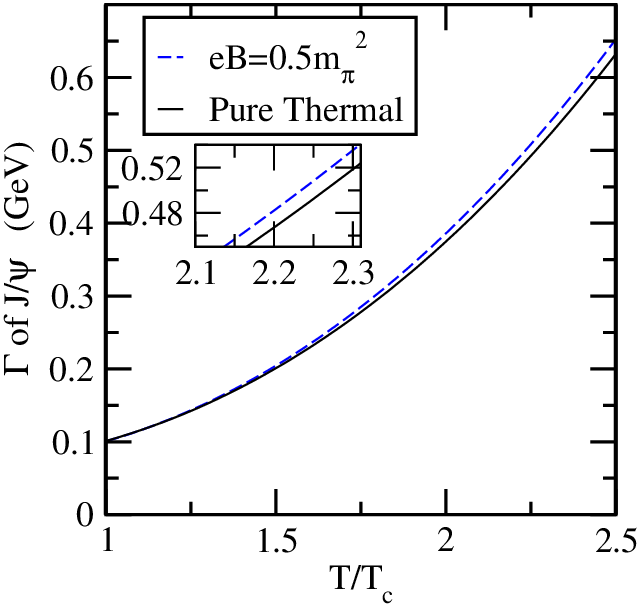}
\end{tabular}
\caption{Variation of the thermal widths with the temperature for 
$J/\psi$.}
\label{decay_psi}
\end{center}
\end{figure}
\begin{figure}[h]
\begin{center}
\begin{tabular}{c c}
\includegraphics[width=7.5cm,height=7.5cm]{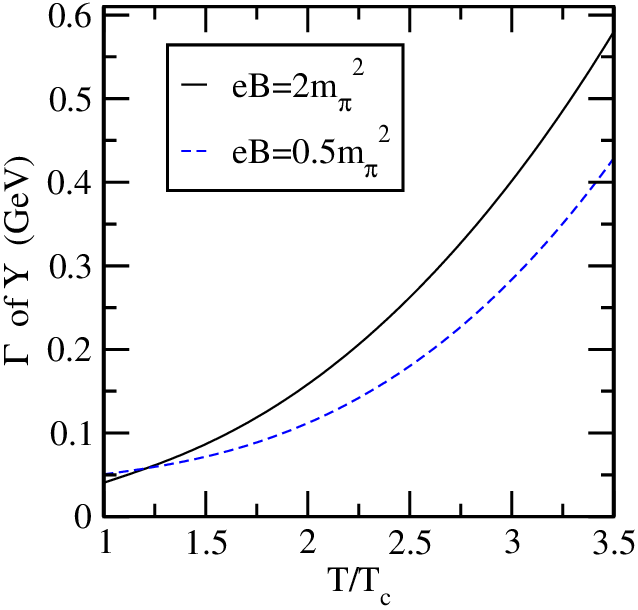}&
~~~~\includegraphics[width=7.5cm,height=7.5cm]{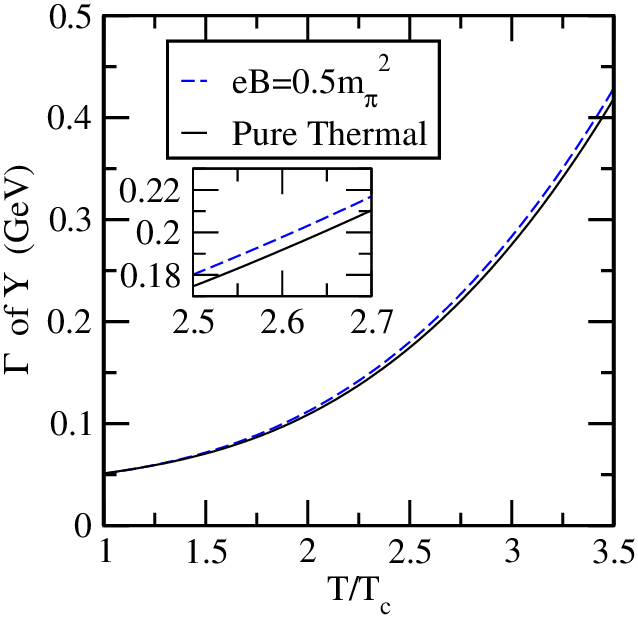}
\end{tabular}
\caption{Variation of the thermal widths with the temperature for 
$\Upsilon$.}
\label{decay_upsilon}
\end{center}
\end{figure}
We have obtained the thermal width numerically and observed that it depend on the temperature as well as the weak 
magnetic field.
To explore the effects of the weak magnetic field on the 
thermal width of heavy quarkonia, we have plotted the 
thermal width of $J/\psi$ and $\Upsilon$ as a function of 
$T/T_c$ for different strengths of magnetic field in 
Fig.\ref{decay_psi} and Fig.\ref{decay_upsilon}, respectively. 
We observed that the thermal widths for $J/\psi$ and 
$\Upsilon$ get increased both with the temperature and 
magnetic field as depicted in the left panels of 
Fig.\ref{decay_psi} and Fig.\ref{decay_upsilon}. We can 
understood this finding in terms of the increase of the 
imaginary-part of the potential, the magnitude of which 
gets enhanced both with temperature and magnetic field. 
We also made a comparison of thermal width in presence 
of weak magnetic field with its counter part in absence 
of magnetic field in the right panels of 
Fig.\ref{decay_psi} and Fig.\ref{decay_upsilon}, where we 
found that the decay widths for $J/\Psi$ and $\Upsilon$ 
get increased in the presence of magnetic field as compared 
to the pure thermal case.

\subsection{Dissociation of quarkonia}
In the previous subsections, we have obtained the binding energies 
and thermal widths of heavy quarkonia, $J/\psi$ and $\Upsilon$. 
Now we will study the quasi-free dissociation of heavy quarkonia 
in a thermal QCD medium and see how the dissociation temperatures 
of quarkonia are affected in the presence of weak magnetic field. 
For that purpose we use the criterion on the width of the resonance 
($\Gamma$): $\Gamma \ge 2 ~{\rm{BE}}$ \cite{Mocsy:PRL99'2007} 
(where ${\rm{BE}}$ is the binding energy of the heavy quarkonia) 
to estimate the dissociation temperature for $J/\psi$ 
and $\Upsilon$.
\begin{table}[H]
\begin{center}
\begin{tabular}{|c|c|c|}
\hline
&\multicolumn{2}{|c|}{Dissociation Temperatures $T_d$ in $T_c$} \\
\hline
State & $J/\psi$ & $\Upsilon$ \\ 
\hline 
Pure Thermal ($eB=0$)& 1.80 & 3.50 \\ 
\hline 
$eB=0.5m_\pi^2$ & 1.74 & 3.43 \\ 
\hline
\end{tabular}
\end{center}
\caption{Dissociation temperatures in absence and 
presence of weak magnetic field.} 
\label{table_diss}
\end{table}
We have obtained the dissociation temperatures of 
$J/\Psi$ and $\Upsilon$ in the absence and presence of 
weak magnetic field in Table.\ref{table_diss}, and 
observed that the dissociation temperatures become 
slightly lower in the presence of weak magnetic field.
{\em For example}, with $eB = 0m_\pi^2$ the $J/\psi$ and 
$\Upsilon$ are dissociated at $1.80T_c$ and $3.50T_c$, 
respectively whereas with $eB = 0.5m_\pi^2$ the $J/\psi$ 
and $\Upsilon$ are dissociated at $1.74T_c$ and $3.43T_c$.
This observation leads to the slightly early dissociation 
of heavy quarkonia in the presence of the weak magnetic field.

\section{Conclusions}
In the present theoretical study, we have explored the 
effects of weak magnetic field on the dissociation of 
quarkonia in a thermal QCD by calculating the 
complex heavy quark potential perturbatively
in the aforesaid medium. For that purpose, we first evaluate 
the gluon self-energy in a similar environment 
using the imaginary-time formalism.
Furthermore, we have revisited the general structure of 
gluon self-energy tensor in the presence of weak magnetic field in 
thermal medium and obtained the relevant structure functions, that 
in turn give rise to the real and imaginary parts of the
resummed gluon propagator, which give the real and imaginary 
parts of the dielectric permittivity. To include the medium 
modification to the non-perturbative part of the vacuum heavy 
quark potential, we have included a non-perturbative term in 
the resummed gluon propagator induced by the dimension two 
gluon condensate besides the usual hard thermal
loop resummed contribution. Thus, the real and imaginary 
parts of the dielectric permittivity will be used to evaluate 
the real and imaginary parts of the complex heavy quark 
potential. We have studied the effects of weak magnetic field
on the real and imaginary parts of the potential. We have 
found that the real-part of the potential is found to be more 
screened on increasing the value of temperature and magnetic field 
both. In addition to this, we have observed that the real-part
gets slightly more screened in the presence of weak magnetic 
field as compared to its counter part in the 
absence of magnetic field. 
On the other hand,  
the magnitude of the imaginary-part of the potential gets 
increased with the value of both temperature and magnetic field,
and its magnitude also gets increased in the presence 
of weak magnetic field as compared to pure thermal case.
The real part of the potential is used in the Schr\"{o}dinger 
equation to obtain the binding energy of heavy quarkonia, 
whereas the imaginary part is used to calculate the 
thermal width. We observed that the binding energies of 
$J/\Psi$ and $\Upsilon$ are found to decrease with the 
temperature and magnetic field both, we can attribute this 
findings in terms of the increasing of screening of the
real-part of the potential. We also observed that the binding energy of $J/\Psi$
and $\Upsilon$ in the presence of magnetic field are 
smaller as compared to the one in the pure thermal case. The 
increase in the magnitude of the imaginary-part of the 
potential will leads to the increase of decay width with 
temperature and magnetic field both. The thermal width for 
$J/\Psi$ and $\Upsilon$ get increased in presence of magnetic 
field as compared to pure thermal case. With the observations 
of binding energy and decay width in hands, we have 
finally studied the dissociation of quarkonia in the presence 
of weak magnetic field. The dissociation temperatures for 
$J/\Psi$ and $\Upsilon$ become slightly lower in the the presence 
of weak magnetic field. {\em For example}, with $eB = 0 m_\pi^2$ 
the $J/\psi$ and $\Upsilon$ 
are dissociated at $1.80T_c$ and $3.50T_c$, respectively whereas 
with $eB = 0.5m_\pi^2$ the $J/\psi$ and $\Upsilon$ are 
dissociated at $1.74T_c$ and $3.43T_c$. This observation leads 
to the slightly early dissociation of quarkonia because of the 
presence of a weak magnetic field.

\section*{Acknowledgements}
One of the author BKP is thankful to the CSIR (Grant No.03 (1407)/17/EMR-II), 
Government of India for the financial assistance.
\appendix
\appendixpage
\addappheadtotoc
In the following appendices we have shown the explicit 
calculations of form factors $b_0(Q)$ and $b_2(Q)$.
\begin{appendices}
\section{Calculation of the form factor $b_0(Q)$}
\label{b_0}
In this appendix, we will use the imaginary-time
formalism to calculate the form factor $b_0$, which 
is given by 
\begin{eqnarray}
b_0(Q)&=&\sum_f \frac{i2g^2}{\bar{u}^2}\int\frac{d^4K}{(2\pi)^4}\frac
{\left[2k_0^2-K^2+m_f^2\right]}
{(K^2-m^2_f)(P^2-m_f^2)},\nonumber\\
&=&-N_f \frac{2g^2}{\bar{u}^2}\int\frac{d^3k}{(2\pi)^3}T\sum_n\frac
{\left[K^2+2k^2\right]}
{(K^2-m^2_f)(P^2-m_f^2)},\nonumber\\
&=&-N_f \frac{2g^2}{\bar{u}^2}[I_1(Q)+I_2(Q)],
\label{formfactor1_b0}
\end{eqnarray}
where we have neglected $m_f$ in numerator in the Hard 
Thermal Loop (HTL) approximation and $\int\frac{d^4K}{(2\pi)^4} 
\rightarrow iT\int\frac{d^3k}{(2\pi)^3}\sum_n$, 
the $I_1$ and $I_2$ are given as 
\begin{eqnarray}
I_1(Q)&=&\int\frac{d^3k}{(2\pi)^3}T\sum_n\frac
{K^2}{(K^2-m^2_f)(P^2-m_f^2)},\\
I_2(Q)&=&\int\frac{d^3k}{(2\pi)^3}T\sum_n\frac
{2k^2}{(K^2-m^2_f)(P^2-m_f^2)}.
\end{eqnarray}
Now we substitute $k_0=i\omega_n$, $q_0=i\omega$, $E_1=\sqrt{k^2+m^2_f}$
and $E_2=\sqrt{(k-q)^2+m^2_f}$, and then perform the frequency sum,
which gives $I_1$ as 
\begin{eqnarray}
I_1(Q)&=&-\int\frac{d^3k}{(2\pi)^3}T\sum_n\frac
{1}{(\omega_n^2+E_1^2)},\nonumber\\
&=&-\int\frac{d^3k}{(2\pi)^3}\frac{1}{2E_1}[1-2n_F(E_1)],
\end{eqnarray}
where the first term is the non-leading term in $T$, thus
retaining only the leading term in $T$, the $I_1$ becomes  
\begin{eqnarray}
I_1(Q)&=&\int\frac{d^3k}{(2\pi)^3}\frac{n_F(E_1)}{E_1},
\end{eqnarray}
now taking $I_2$, which becomes
\begin{eqnarray}
I_2(Q)&=&2\int\frac{d^3k}{(2\pi)^3}k^2~T\sum_n\frac
{1}{(\omega_n^2+E_1^2)[{(\omega_n-\omega)}^2+E_2^2]},\nonumber\\
&=&-\int\frac{d^3k}{(2\pi)^3}\left[\frac{n_F(E_1)}{E_1}
+q\cos\theta\frac{dn_F(E_1)}{dk}\frac{1}{i\omega-
q\cos\theta}\right].
\end{eqnarray}
Substituting $I_1$ and $I_2$ in Eq.\eqref{formfactor1_b0}, 
the form factor $b_0$ becomes
\begin{eqnarray}
b_0(q_0,q)=-N_f \frac{2g^2}{\bar{u}^2}\int\frac{d^3k}{(2\pi)^3}
\frac{dn_F(E_1)}{dk}\left(1-\frac{q_0}{q_0-q\cos\theta}\right),
\end{eqnarray}
where we have again resubstituted $q_0=i\omega$. Now we will 
evaluate the real and imaginary parts of the form factor 
$b_0$. 

The real-part of $b_0$ in the static limit is given by
\begin{eqnarray}
\rm{Re}~b_0(q_0=0)&=&-N_f \frac{g^2}{\pi^2}\int k^2 dk
\frac{dn_F(E_1)}{dk},\nonumber\\
&=&N_f\frac{g^2T^2}{6}.
\end{eqnarray}
On the other hand, for the evaluation of the 
imaginary part of $b_0$ we will us the following 
identity
\begin{eqnarray}
{\rm Im}~b_0(q_0,q)=\frac{1}{2i}\lim_{\eta\rightarrow 0}
\left[b(q_0+i\eta,q)-b(q_0-i\eta,q)\right],
\label{identity1}
\end{eqnarray} 
along with the following expression
\begin{eqnarray}
\frac{1}{2i}\left(\frac{1}{q_0+\sum_j E_j+i\eta}-\frac{1}
{q_0+\sum_j E_j-i\eta}\right)=-\pi\delta(q_0+\sum_j E_j).
\label{identity2}
\end{eqnarray}
Thus using the above identities Eq.\eqref{identity1} 
and Eq.\eqref{identity2}, 
the imaginary-part of $b_0$ becomes
\begin{eqnarray}
\rm {Im}~b_0(q_0,q)&=&N_f\frac{2g^2}{\bar{u}^2}
\frac{1}{2i}\lim_{\eta\rightarrow 0}\int \frac{d^3k}{(2\pi)^3}
\frac{dn_F(k)}{dk}
\left(\frac{q\cos\theta}{q_0-q\cos\theta+i\eta}
-\frac{q\cos\theta}{q_0-q\cos\theta-i\eta}\right),\nonumber\\
&=&-N_f\frac{\pi g^2}{2\pi^2\bar{u}^2}
\frac{q_0}{q}\int k^2~dk\frac{dn_F(k)}{dk},
\end{eqnarray}
which in the static limit takes the simplified 
form 
\begin{eqnarray}
\left[\frac{{\rm Im}~b_0(q_0,q)}{q_0}\right]_
{q_0=0}=\frac{g^2T^2N_f}{6}\frac{\pi}{2q}.
\end{eqnarray} 

\section{Calculation of the form factor $b_2(Q)$}
\label{b_2}
Similar to the form factor $b_0$, here we will solve the 
form factor $b_2$, which is given by
\begin{eqnarray}
b_2(Q)&=&\sum_f \frac{i2g^2(q_fB)^2}{\bar{u}^2}\left[\int\frac
{d^4K}{(2\pi)
^4}\left\lbrace\frac{\left(2k_0^2-K_\parallel^2+m_f^2\right)}
{(K^2-m^2_f)^2(P^2-m_f^2)^2}
-\frac{\left(8k_0^2K_\perp^2\right)}{(K^2-m^2_f)^4(P^2-m_f^2)}\right
\rbrace\right],\nonumber\\
&=&-\sum_f \frac{2g^2(q_fB)^2}{\bar{u}^2}\int\frac{d^3k}{(2\pi)^3}
T\sum_n\left\lbrace\frac{K^2+k^2(1+\cos^2\theta)+m_f^2)}
{(K^2-m^2_f)^2(P^2-m_f^2)^2}\right.\nonumber\\&&\left.-
\frac{8(k^4+k^2K^2)(1-\cos^2\theta)}
{(K^2-m^2_f)^4(P^2-m_f^2)}\right\rbrace,
\end{eqnarray}
where we have used the spherical polar coordinate 
system for $k=(k\sin\theta\sin\phi,k\sin\theta
\cos\phi,k\cos\theta)$. In order
to solve the form factor $b_2$, we will use the method 
as shown in~\cite{karmakar:EPJC79'2019} , which gives 
\begin{eqnarray}
b_2(Q)&=&-\sum_f\frac{2g^2q_f^2B^2}{\bar{u}^2}\left[
\left\lbrace\frac{\partial}{\partial(m_f^2)}+\frac{5}{6}
m_f^2\frac
{\partial^2}{\partial^2(m_f^2)}\right\rbrace
\int\frac{d^3k}{(2\pi)^3}T\sum_n\frac{1}{(K^2-m_f^2)
(P^2-m_f^2)}\right.\nonumber\\&&\left.-\left\lbrace\frac
{\partial}{\partial(m_f^2)}+\frac{m_f^2}{2}\frac
{\partial^2}{\partial^2(m_f^2)}\right\rbrace
\int\frac{d^3k}{(2\pi)^3}T\sum_n\frac{\cos^2\theta}
{(K^2-m_f^2)(P^2-m_f^2)}
\right],
\end{eqnarray}
and now we will perform the following frequency sum
\begin{eqnarray}
T\sum_n\frac{1}{(\omega_n^2+E_1^2)[(\omega_n-\omega)^2+E_2^2]}
&=&\frac{[1-n_F(E_1)-n_F(E_2)]}{4E_1E_2}\left\lbrace\frac{1}
{i\omega+E_1+E_2}-\frac{1}{i\omega-E_1-E_2}\right\rbrace\nonumber\\
&&+\frac{[n_F(E_1)-n_F(E_2)]}{4E_1E_2}\left\lbrace\frac{1}
{i\omega+E_1-E_2}-\frac{1}{i\omega-E_1+E_2}\right\rbrace.
\end{eqnarray}
Thus, after simplification the form factor $b_2$ becomes
\begin{eqnarray}
b_2(q_0,q)&=&\sum_f\frac{2g^2q_f^2B^2}{\bar{u}^2}\left\lbrace\
\left(\frac{\partial^2}{\partial^2(m_f^2)}+\frac{5}{6}
m_f^2\frac{\partial^3}{\partial^3(m_f^2)}\right)
\int\frac{d^3k}{(2\pi)^3}\frac{n_F(E_1)}{E_1}\left
(\frac{q_0}{q_0-q\cos\theta}-1\right)\right.\nonumber\\&&+\left.
\left(\frac{\partial}{\partial(m_f^2)}+\frac{5}{6}
m_f^2\frac{\partial^2}{\partial^2(m_f^2)}\right)
\int\frac{d^3k}{(2\pi)^3}\frac{n_F(E_1)}{2E_1^3}
\left(\frac{q_0}{q_0-q\cos\theta}\right)
\right.\nonumber\\&&-\left.
\left(\frac{\partial^2}{\partial^2(m_f^2)}+\frac{
m_f^2}{2}\frac{\partial^3}{\partial^3(m_f^2)}\right)
\int\frac{d^3k}{(2\pi)^3}\frac{n_F(E_1)}{E_1}\cos^2\theta
\left(\frac{q_0}{q_0-q\cos\theta}-1\right)
\right.\nonumber\\&&-\left.
\left(\frac{\partial}{\partial(m_f^2)}+\frac{
m_f^2}{2}\frac{\partial^2}{\partial^2(m_f^2)}\right)
\int\frac{d^3k}{(2\pi)^3}\frac{n_F(E_1)}{2E_1^3}\cos^2\theta
\left(\frac{q_0}{q_0-q\cos\theta}\right)
\right\rbrace.
\end{eqnarray}
Thus, the real part of $b_2$ in the static 
limit is obtained as
~\cite{karmakar:EPJC79'2019}
\begin{eqnarray}
{\rm Re}~b_2(q_0=0)=\sum_f\frac{g^2}{12\pi^2 T^2}(q_fB)^2 
\sum_{l=1}^{\infty}(-1)^{l+1}l^2K_0(\frac{m_fl}{T}).
\end{eqnarray}
Now we will evaluate the imaginary part of form factor $b_2$,
for that we write $b_2$ as 
\begin{eqnarray}
b_2(q_0,q)=\sum_f\frac{2g^2q_f^2B^2}{\bar{u}^2}[I_3(q_0,q)+
I_4(q_0,q)+I_5(q_0,q)+I_6(q_0,q)],
\label{sum}
\end{eqnarray}
where we have defined the following functions:
\begin{eqnarray}
I_3(q_0,q)&=&
\left(\frac{\partial^2}{\partial^2(m_f^2)}+\frac{5}{6}
m_f^2\frac{\partial^3}{\partial^3(m_f^2)}\right)
\int\frac{d^3k}{(2\pi)^3}\frac{n_F(E_1)}{E_1}\left
(\frac{q\cos\theta}{q_0-q\cos\theta}\right),\\
I_4(q_0,q)&=&
\left(\frac{\partial}{\partial(m_f^2)}+\frac{5}{6}
m_f^2\frac{\partial^2}{\partial^2(m_f^2)}\right)
\int\frac{d^3k}{(2\pi)^3}\frac{n_F(E_1)}{2E_1^3}
\left(\frac{q_0}{q_0-q\cos\theta}\right),\\
I_5(q_0,q)&=&-
\left(\frac{\partial^2}{\partial^2(m_f^2)}+\frac{
m_f^2}{2}\frac{\partial^3}{\partial^3(m_f^2)}\right)
\int\frac{d^3k}{(2\pi)^3}\frac{n_F(E_1)}{E_1}\cos^2\theta
\left(\frac{q\cos\theta}{q_0-q\cos\theta}\right),\\
I_6(q_0,q)&=&-
\left(\frac{\partial}{\partial(m_f^2)}+\frac{
m_f^2}{2}\frac{\partial^2}{\partial^2(m_f^2)}\right)
\int\frac{d^3k}{(2\pi)^3}\frac{n_F(E_1)}{2E_1^3}\cos^2\theta
\left(\frac{q_0}{q_0-q\cos\theta}\right),
\end{eqnarray}
Now we will evaluate the imaginary parts of all the
above four terms one by one using the identities
Eq.\eqref{identity1} and Eq.\eqref{identity2}, first 
we start with $I_3(q_0,q)$
\begin{eqnarray}
{\rm Im} I_3(q_0,q)&=&X_3(m_f)
\frac{1}{2i}\lim_{\eta\rightarrow 0}\left[
\int\frac{d^3k}{(2\pi)^3}
\frac{n_F(E_1)}{E_1}
\left(\frac{q\cos\theta}{q_0-q\cos\theta+i\eta}
-\frac{q\cos\theta}{q_0-q\cos\theta-i\eta}\right)\right],
\nonumber\\\label{function_3}
\end{eqnarray}
where $X_3(m_f)=\left(\frac{\partial^2}{\partial^2
(m_f^2)}+\frac{5}{6}
m_f^2\frac{\partial^3}{\partial^3(m_f^2)}\right)$,
now the Eq.\eqref{function_3} in the static limit becomes
\begin{eqnarray}
\left[\frac{{\rm Im}~I_3(q_0,q)}{q_0}\right]_{q_0=0}&=
&-\frac{1}{4\pi q}X_3(m_f)
\int k^2~dk\frac{n_F(E_1)}{E_1},\nonumber\\
&=&-\frac{1}{4\pi q}X_3(m_f)
\sum_{l=1}^{\infty}\frac{m_f^2}{2}\left[K_2
(\frac{m_fl}{T})-K_0(\frac{m_fl}{T})\right],\nonumber\\
&=&\frac{1}{32\pi q T^2}
\sum_{l=1}^{\infty}(-1)^{l+1}l^2 K_0(\frac{m_fl}{T})
-\frac{1}{192\pi q T^2}
\sum_{l=1}^{\infty}(-1)^{l+1}l^2 K_2(\frac{m_fl}{T}),
\nonumber\\\label{i3}
\end{eqnarray}
where $K_0$ and $K_2$ are the modified Bessel functions 
of second kind. Now we take $I_4(q_0,q)$
\begin{eqnarray}
{\rm Im} I_4(q_0,q)&=&X_4(m_f)
\frac{1}{2i}\lim_{\eta\rightarrow 0}\left[
\int\frac{d^3k}{(2\pi)^3}
\frac{n_F(E_1)}{2E_1^3}
\left(\frac{q_0}{q_0-q\cos\theta+i\eta}
-\frac{q_0}{q_0-q\cos\theta-i\eta}\right)\right],
\nonumber\\\label{function_4}
\end{eqnarray}
where $X_4(m_f)=\left(\frac{\partial}{\partial
(m_f^2)}+\frac{5}{6}
m_f^2\frac{\partial^2}{\partial^2(m_f^2)}\right)$
and the Eq.\eqref{function_4}, takes the following 
form in the static limit
\begin{eqnarray}
\left[\frac{{\rm Im}~I_4(q_0,q)}{q_0}\right]_{q_0=0}&=
&-\frac{1}{8\pi q}X_4(m_f)
\int k^2~dk\frac{n_F(E_1)}{E_1^3},\nonumber\\
&=&\frac{1}{16\pi q}X_4(m_f)\left[1+\gamma_E-\frac
{\pi m_f}{4T}+\log\frac{m_f}{\pi T}\right],\nonumber\\
&=&\frac{1}{1536\pi q}\frac{(8T-7\pi m_f)}{m_f^2T}. 
\nonumber\\\label{i4}
\end{eqnarray}
Similarly
the imaginary part of $I_5(q_0,q)$ and $I_6(q_0,q)$
\begin{eqnarray}
{\rm Im} I_5(q_0,q)&=&-X_5(m_f)
\frac{1}{2i}\lim_{\eta\rightarrow 0}\left[
\int\frac{d^3k}{(2\pi)^3}
\frac{n_F(E_1)}{E_1}
\left(\frac{q\cos^3\theta}{q_0-q\cos\theta+i\eta}
-\frac{q\cos^3\theta}{q_0-q\cos\theta-i\eta}\right)
\right],
\nonumber\\\label{function_5}
\end{eqnarray}
where $X_5(m_f)=\left(\frac{\partial^2}{\partial^2
(m_f^2)}+\frac{
m_f^2}{2}\frac{\partial^3}{\partial^3(m_f^2)}\right)$,
and the Eq.\eqref{function_5} vanishes in the static limit
\begin{eqnarray}
\left[\frac{{\rm Im}~I_5(q_0,q)}{q_0}\right]_{q_0=0}&=
&0,
\label{i5}
\end{eqnarray}
\begin{eqnarray}
{\rm Im} I_6(q_0,q)&=&-X_6(m_f)
\frac{1}{2i}\lim_{\eta\rightarrow 0}\left[
\int\frac{d^3k}{(2\pi)^3}
\frac{n_F(E_1)}{2E_1^3}
\left(\frac{q_0\cos^2\theta}{q_0-q\cos\theta+i\eta}
-\frac{q_0\cos^2\theta}{q_0-q\cos\theta-i\eta}\right)
\right],
\nonumber\\\label{function_6}
\end{eqnarray}
where $X_6(m_f)=\left(\frac{\partial}{\partial
(m_f^2)}+\frac{
m_f^2}{2}\frac{\partial^2}{\partial^2(m_f^2)}\right)$,
the Eq.\eqref{function_6} also vanishes in the 
static limit
\begin{eqnarray}
\left[\frac{{\rm Im}~I_6(q_0,q)}{q_0}\right]_{q_0=0}&=
&0.
\label{i6}
\end{eqnarray}
Finally, we substitute Eq.\eqref{i3}, Eq.\eqref{i4}, Eq.\eqref{i5} and 
Eq.\eqref{i6} in Eq.\eqref{sum}, to evaluate the imaginary part 
of $b_2(q_0,q)$, which in the static limit can be written as  
\begin{eqnarray}
\left[\frac{{\rm Im}~b_2(q_0,q)}{q_0}\right]_
{q_0=0}&=&\frac{1}{q}\left[\sum_f\frac{g^2(q_fB)^2}{16\pi T^2}
\sum_{l=1}^\infty(-1)^{l+1}l^2K_0\left(\frac{m_f l}{T}\right)
\right. \nonumber\\ &&\left. 
-\sum_f\frac{g^2(q_fB)^2}{96\pi T^2}
\sum_{l=1}^\infty(-1)^{l+1}l^2K_2\left(\frac{m_f l}{T}\right)
\right. \nonumber\\ &&\left.
+\sum_f\frac{g^2(q_fB)^2}{768\pi}\frac{(8T-7\pi m_f)}{m_f^2 T}
\right].
\end{eqnarray}
\end{appendices}


\end{document}